\documentclass[twocolumn]{aastex62}

\usepackage{array}
\usepackage{amsmath}
\usepackage[counterclockwise, figuresleft]{rotating}
\usepackage{txfonts}
\usepackage{booktabs}
\usepackage{multirow}
\usepackage[caption=false]{subfig}
\usepackage{longtable}
\usepackage{lineno}
\usepackage{hyperref}

\newcommand{\mstar}{M_\star}
\newcommand{\rstar}{R_\star}

\newcommand{\teq}{T_\mathrm{eq}}

\newcommand{\teff}{T_\mathrm{eff}}

\def\ms{\hbox{\,m\,s$^{-1}$}}         
\def\m2s2{\hbox{\,m$^{2}$\,s$^{-2}$}} 
\def\kms{\hbox{\,km\,s$^{-1}$}}       
\def\gcm{\hbox{\,g\,cm$^{-1}$}}       
\def\vsini{\hbox{$v$\,sin\,$i$}}      
\def\sini{\hbox{sin\,$i$}}      

\def\mp{M_{\rm p}}

\def\Porb{P_{\rm orb}}
\def\Prot{P_{\rm rot}}

\begin{document}


\title{The HD~60779 Planetary System: A Transiting Sub-Neptune on a 30-day Orbit and a More Massive Outer World}


\author[0000-0003-0741-7661]{Victoria DiTomasso}\thanks{Corresponding author\\victoria.ditomasso@cfa.harvard.edu}

\author[0000-0002-9003-484X]{David Charbonneau}
\affiliation{Center for Astrophysics | Harvard and Smithsonian, 60 Garden Street, Cambridge, MA 02138, USA}

\author[0000-0001-7246-5438]{Andrew Vanderburg}
\affiliation{Center for Astrophysics | Harvard and Smithsonian, 60 Garden Street, Cambridge, MA 02138, USA}
\affiliation{Kavli Institute for Astrophysics and Space Research, Massachusetts Institute of Technology, Cambridge, MA 02139, USA}

\author[0000-0003-3204-8183]{Mercedes López-Morales}
\affiliation{Space Telescope Science Institute, 3700 San Martin Drive, Baltimore, MD 21218, USA}

\author[0000-0003-2527-1475]{Shreyas Vissapragada}
\affiliation{Carnegie Science Observatories, 813 Santa Barbara Street, Pasadena, CA 91101, USA}

\author[0000-0001-7254-4363]{Annelies Mortier}
\affiliation{School of Physics \& Astronomy, University of Birmingham, Edgbaston, Birmingham, B15 2TT, UK}

\author[0000-0001-8749-1962]{Thomas G. Wilson}
\affiliation{Department of Physics, University of Warwick, Gibbet Hill Road, Conventry, CV4 7LS, UK}

\author[0000-0001-5637-6144]{Elyse Incha}
\affiliation{Department of Astronomy, University of Wisconsin-Madison, 475 N. Charter St., Madison, WI 53703, USA}

\author[0000-0002-8863-7828]{Andrew Collier Cameron}
\affiliation{Centre for Exoplanet Science, SUPA, School of Physics and Astronomy, University of St Andrews, St Andrews KY16 9SS, UK}

\author[0000-0002-6492-2085]{Luca Malavolta}
\affiliation{Dipartimento di Fisica e Astronomia “Galileo Galilei”, Universitá di Padova, Vicolo del l’Osservatorio 3, I-35122 Padova, Italy}

\author[0000-0003-1605-5666]{Lars A. Buchhave}
\affiliation{DTU Space, National Space Institute, Technical University of Denmark, Elektrovej 328, DK-2800 Kgs. Lyngby, Denmark}

\author[0000-0001-9911-7388]{David W. Latham}
\affiliation{Center for Astrophysics | Harvard and Smithsonian, 60 Garden Street, Cambridge, MA 02138, USA}


\author[0000-0002-4445-1845]{Matteo Pinamonti}
\affiliation{INAF - Osservatorio Astrofisico di Torino, Via Osservatorio 20, I-10025 Pino Torinese, Italy}


\author[0009-0008-5145-0446]{Stephanie Striegel}
\affiliation{SETI Institute, Mountain View, CA 94043 USA}
\affiliation{NASA Ames Research Center, Moffett Field, CA 94035 USA}


\author[0000-0002-9113-7162]{Michael Fausnaugh}
\affiliation{Kavli Institute for Astrophysics and Space Research, Massachusetts Institute of Technology, Cambridge, MA 02139, USA}

\author[0000-0002-0514-5538]{Luke Bouma}
\affiliation{Carnegie Science Observatories, 813 Santa Barbara Street, Pasadena, CA 91101, USA}

\author{Ben Falk}
\affiliation{Space Telescope Science Institute, 3700 San Martin Drive, Baltimore, MD 21218, USA}


\author[0000-0003-2822-616X]{Robert Aloisi}
\affiliation{Department of Astronomy, University of Wisconsin-Madison, 475 N. Charter St., Madison, WI 53703, USA}

\author[0000-0002-9332-2011]{Xavier Dumusque}
\affiliation{Département d’astronomie de l’Université de Genève, Chemin Pegasi 51, 1290 Versoix, Switzerland}

\author[0000-0002-1715-6939]{A. Anna John}
\affiliation{School of Physics \& Astronomy, University of Birmingham, Edgbaston, Birmingham, B15 2TT, UK}

\author[0000-0002-8122-2240]{Ben S. Lakeland}
\affiliation{School of Physics \& Astronomy, University of Birmingham, Edgbaston, Birmingham, B15 2TT, UK}

\author[0000-0002-4272-4272]{A. F. Martínez Fiorenzano}
\affiliation{Fundación Galileo Galilei - INAF, Rambla José Ana Fernandez Pérez 7, E-38712 Breña Baja, Tenerife, Spain}

\author[0000-0001-9390-0988]{Luca Naponiello}
\affiliation{INAF - Osservatorio Astrofisico di Torino, Via Osservatorio 20, I-10025 Pino Torinese, Italy}

\author[0000-0003-1360-4404]{Belinda Nicholson}
\affiliation{Centre for Astrophysics, University of Southern Queensland, West St., Toowoomba, Queensland, Australia, 4350}

\author[0000-0002-1533-9029]{Emily K. Pass}
\affiliation{Kavli Institute for Astrophysics and Space Research, Massachusetts Institute of Technology, Cambridge, MA 02139, USA}

\author[0000-0002-9815-773X]{Francesco Alfonso Pepe}
\affiliation{Département d’astronomie de l’Université de Genève, Chemin Pegasi 51, 1290 Versoix, Switzerland}

\author[0000-0002-0594-7805]{Federica Rescigno}
\affiliation{Department of Astrophysics, University of Birmingham, Edgbaston, Birmingham, B15 2TT, UK}

\author[0000-0002-7504-365X]{Alessandro	Sozzetti}
\affiliation{INAF - Osservatorio Astrofisico di Torino, Via Osservatorio 20, I-10025 Pino Torinese, Italy}

\author[0009-0001-0867-9711]{Daisy A. Turner}
\affiliation{School of Physics \& Astronomy, University of Birmingham, Edgbaston, Birmingham, B15 2TT, UK}


\author{Saul A.~Rappaport}
\affiliation{Kavli Institute for Astrophysics and Space Research, Massachusetts Institute of Technology, Cambridge, MA 02139, USA}

\author{Mark Omohundro}%
\affiliation{Citizen Scientist, c/o Zooniverse, Department of Physics, University of Oxford, Denys Wilkinson Building, Keble Road, Oxford, OX1 3RH, UK}

\author[0000-0003-0501-2636]{Brian P. Powell}
\affiliation{NASA Goddard Space Flight Center, 8800 Greenbelt Road, Greenbelt, MD 20771, USA}
\author[0000-0002-5665-1879]{Robert Gagliano}%
\affiliation{Amateur Astronomer, Glendale, Arizona, USA}

\author[0000-0003-3988-3245]{Thomas L. Jacobs}%
\affiliation{Amateur Astronomer, Missouri City, Texas 77459 USA}

\author[0000-0001-9786-1031]{Veselin~B.~Kostov}
\affiliation{NASA Goddard Space Flight Center, 8800 Greenbelt Road, Greenbelt, MD 20771, USA}
\affiliation{SETI Institute, Mountain View, CA 94043 USA}
\affiliation{GSFC Sellers Exoplanet Environments Collaboration}

\author[0000-0002-2607-138X]{Martti H. Kristiansen}%
\affiliation{Brorfelde Observatory, Observator Gyldenkernes Vej 7, DK-4340 T\o{}ll\o{}se, Denmark}

\author[0000-0002-8527-2114]{Daryll M. LaCourse}%
\affiliation{Amateur Astronomer, 7507 52nd Pl NE, Marysville, WA, 98270, USA}

\author[0000-0002-5034-0949]{Allan R.~Schmitt}
\affiliation{Citizen Scientist, 616 W. 5. St., Apt. 101, Minneapolis, MN 55419, USA}

\author{Hans Martin Schwengeler}%
\affiliation{Citizen Scientist, Planet Hunter, Bottmingen, Switzerland}

\author{Ivan A. Terentev}  
\affiliation{Citizen Scientist, Planet Hunter, Petrozavodsk, Russia}



\begin{abstract}

{
We present the discovery of the planetary system orbiting the bright ($V=7.2$), nearby (35\,pc), Sun-like ($\mstar=1.050\pm 0.044$\,M$_\odot$, $R_\star=1.129\pm0.013$\,R$_\odot$) star, HD~60779.
We report two TESS transits and a subsequent CHEOPS transit of HD~60779~b, a sub-Neptune ($3.250^{+0.100}_{-0.098} $\,R$_\oplus$) planet on a $29.986175^{+0.000030}_{-0.000033}$\,d orbit. Additionally, 286 HARPS-N radial velocity measurements reveal the mass of planet b ($14.7^{+1.1}_{-1.0} $\,M$_\oplus$) and the presence of an outer planet, HD~60779~c ($\Porb =104.25^{+0.30}_{-0.29}$\,d, $\mp\sini=27.7 \pm 1.6 $\,M$_\oplus$). Both of these planets' orbits are consistent with being circular, suggesting that they have a dynamically quiet history. The data are not sufficient to determine whether planet c transits. HD~60779’s uniquely high systemic radial velocity ($129.75\pm0.12$\,\kms) allows its Lyman-alpha emission to avoid absorption by the interstellar medium, making it a prime candidate for probing atmospheric escape from HD~60779~b. HD~60779 is also the third brightest host of a $\Porb>25$\,d sub-Neptune with a measured mass and radius, distinguishing it in terms of accessibility to spectroscopic characterization.
}

\smallskip
\smallskip
\smallskip
\smallskip
\smallskip

\end{abstract}

\section{Introduction}\label{sect:intro}

The Kepler space mission \citep{borucki_kepler_2010} revealed that planets between the size of Earth and Neptune are the most common in the galaxy, and they naturally fall into one of two populations –- the super-Earth population ($1$\,R$_\oplus<$ R$_p<1.8$\,R$_\oplus$) and the sub-Neptune population ($1.8$\,R$_\oplus<$ R$_p<4$\,R$_\oplus$), divided by the radius valley \citep{fulton_california-kepler_2017}.
The bulk composition and internal structure of sub-Neptunes is currently unknown: one picture holds that they are terrestrial cores enveloped in a modest envelope of primordial H/He gas \citep[e.g.][]{owen_kepler_2013}, whereas a dissenting view is that they are largely ``waterworlds" with substantial volatile layers \citep{luque_density_2022}.
These competing scenarios tie into theories of their formation and atmospheric evolution.
In the first scenario, both super-Earths and sub-Neptunes formed as volatile-poor rocky cores. Sub-Neptunes, however, may have had larger core masses or longer timescales for envelope accretion \citep{lee_primordial_2021}, or else both populations may have started with envelopes, with only the sub-Neptunes able to retain them against mass-loss processes such as photoevaporation \citep{lopez_understanding_2014} or core-powered mass loss \citep{gupta_sculpting_2019}.
The ``waterworld" scenario attributes the origin of the radius valley to conditions during formation. \citet{venturini_nature_2020} proposed that a bimodality in planetary core mass and composition—arising from the location of formation relative to the water ice line—could produce a radius valley even in the absence of atmospheric escape.
Advances in telescope instrumentation and data processing have launched us into an era of follow-up observations that could help answer these questions.

Taking full advantage of these state-of-the-art telescopes and techniques, however, requires the target planet to meet a variety of criteria: it must transit its host star, and that host star must be bright enough for high-resolution spectroscopic observations. While the Transiting Exoplanet Survey Satellite (TESS) is finding thousands of nearby planets that lend themselves to follow-up observations \citep{guerrero_tess_2021}, it is primarily discovering planets on short orbital periods ($<13$\,d, half the length of a TESS sector). While informative, if we seek to understand planet formation and evolution in the Solar System, we will need to characterize planets on longer-period orbits. HD~60779, a bright ($V=7.2$) Sun-like ($\teff=6081$\,K) star, and its accompanying planet(s) present such an opportunity.

We present the detection and confirmation of the HD~60779 planetary system. First, we present the detection of a single transit of HD~60779~b, which brought this target to our attention. We describe our analysis of this transit, and the reconnaisance spectra that encouraged follow-up (Section \ref{sect:transit_dect_prelim_analysis}). Then, we report our radial velocity (RV) follow-up campaign with the High Accuracy Radial velocity Planet Searcher for the Northern hemisphere spectrograph \citep[HARPS-N,][]{cosentino_harps-n_2012,cosentino_harps-n_2014} and describe our analysis. We discuss our preliminary model fits and analysis which led us to characterize this as a two-planet system, and allowed us to constrain the orbital period of the transiting planet (Section \ref{sect:RV}). We describe our transit follow-up efforts, including predicting transit times, as well as the additional transit observations from CHEOPS and TESS (Section \ref{sect:transit_followup}). We then discuss our characterization of the host star (Section \ref{sect:starchar}). Then we present our final fits and parameters for HD~60779~b and c (Section \ref{sect:final_RV_transit_analysis}). Finally, we summarize our conclusions and discuss potential for further characterization of these planets (Section \ref{sect:conclusions}).



\section{Initial Transit Detection of a Single Transit}\label{sect:transit_dect_prelim_analysis}

\subsection{TESS Data and Initial Transit Detection }\label{subsect:tess_data_init_detect}

TESS is a space-based telescope consisting of four wide-field cameras with an effective aperture of 10\,cm, which was designed to perform an all-sky survey, measuring photometric light curves of stars to detect transiting exoplanets \citep{ricker_transiting_2014}. The TESS survey divides the sky into $24^{\circ} \times 96^{\circ}$ strips, and observes each strip over a baseline of 27 days, known as a sector. TESS observed HD~60779 during three 27-day sectors: 7 (Jan-Feb 2019), 34 (Jan-Feb 2021) and 88 (Jan-Feb 2025). All three sectors have 120\,s cadence data, and Sector 88 also has 20\,s cadence data.

A single transit was identified in the TESS data after the release of Sector 34 (prior to the collection of Sector 88), by the Visual Survey Group (VSG). VSG is a collaboration of professional and amateur astronomers who manually search light curves from the Kepler \citep{borucki_kepler_2010}, K2 \citep{howell_k2_2014} and TESS missions with the goal of identifying transits and other notable astrophysical phenomena \citep{kristiansen_visual_2022}. VSG uses \texttt{LcTools}\footnote{https://sites.google.com/a/lctools.net/lctools/home} \citep{schmitt_lctools_2019,schmitt_lctools_2021} to create normalized light curves, find candidate signals via a Box-Fitting Least Squares algorithm, and visually inspect the light curves.

The transit occurred toward the end of Sector 34, centered at JD $= 2459252.9531$. The depth of this transit corresponded to a sub-Neptune-sized planet 
($1.8-4$\,R$_\oplus$). The detection of only one transit in Sector 34 indicated that the orbital period of this planet candidate (hereafter referred to as HD~60779~b) was $>25$\,days. Even at a period of 25d, this planet would have a longer period than 80\% of sub-Neptunes with a measured radius and mass around a Sun-like star (4000\,K$<\teff<$ 6500\,K)\footnote{According to a search of the NASA Exoplanet Archive Planetary System Composite Table on Jun 11 2025, https://exoplanetarchive.ipac.caltech.edu/}. HD~60779 would also be one of the brightest hosts of a long-period sub-Neptune with a measured mass and radius, fainter only than $\nu^2$ Lupi and 24 LMi. The opportunity to detect and characterize such a system warranted additional efforts.

\subsection{Preliminary Transit Analysis}\label{subsect:prelim_transit_analysis}

Using the available TESS light curves, we sought to constrain the expected orbital period of the transiting planet. We used the Pre-search Data Conditioning Simple Aperture Photometry \citep{smith_kepler_2012, stumpe_kepler_2012,stumpe_multiscale_2014} flux and flux uncertainties as reduced by the Science Processing Operations Center pipeline at NASA Ames Research Center and provided on the Mikulski Archive for Space Telescopes (MAST)\footnote{https://mast.stsci.edu/portal/Mashup/Clients/Mast/Portal.html}. We removed all points with NaN flux values or bad-quality flags (DQUALITY $>$ 0). Following the methodology for single transit analysis described in \cite{incha_keplers_2023}, we modeled the shape of the transit with respect to transit duration, $R_p/R_\star$ and impact parameter, assuming a distribution of $e$ and $\omega$ generated via the \texttt{ECCSAMPLES}\footnote{https://github.com/davidkipping/ECCSAMPLES} software \citep{kipping_bayesian_2014}. We then calculated the probability of different orbital periods, considering what periods are ruled out by the non-detection of an additional transit in the same TESS sector and the geometric transit probability. This resulted in a predicted orbital period distribution with a peak at 26.6 days and a median of 35.3 days. We also used the relationship from \cite{yee_characterizing_2008}, which relates orbital period, stellar density, transit duration, ingress/egress duration and transit depth to estimate an orbital period of 27.5 days. These estimates indicated that we would be able to detect this planet after a feasibly-lengthed RV campaign.

\subsection{Recon Spectra}\label{subsect:recon_spec}

The high-resolution FIbre-fed Echelle Spectrograph (FIES) is a high-resolution (R=67000) spectrograph located on the 2.56\,m Nordic Optical Telescope at the Observatorio del Roque de los Muchachos in La Palma, Spain \citep{telting_fies_2014,djupvik_nordic_2009}.
We obtained three FIES spectra of HD~60779, two taken on March 23, 2021 and a third taken on March 25, 2021.
The primary goal of these observations was to evaluate HD~60779's suitability for precise RV follow-up. Indeed, its low $\vsini$ ($3.6 \pm 0.5$\,\kms) indicated that HD~60779 is suitable for RV follow-up.

\section{Radial Velocity Monitoring}\label{sect:RV}

\subsection{HARPS-N Spectroscopy}\label{subsect:RVdata}

We initiated intensive RV observations of HD~60779 to constrain the orbital period of the transiting planet. We obtained 299 spectra of HD~60779 between May 2021 and April 2025 with the HARPS-N spectrograph ($383-693$\,nm), installed on the 3.6\,m Telescopio Nazionale Galileo (TNG) at the Observatorio del Roque de los Muchachos in La Palma, Spain \citep{cosentino_harps-n_2012,cosentino_harps-n_2014}. Observations were collected as part of the HARPS-N Collaboration\footnote{Guaranteed Time of Observations (2021-2023),  A48TAC\_59, PI: Malavolta (2023-2025)}. The initial season of observations, totaling twelve data points, were affected by a HARPS-N guiding issue that resulted in RV offsets for all targets observed at that time (7-12 May 2021), as documented in Lienhard et al 2025 (submitted) and \cite{nicholson_hd152843_2024}. As done in those works, we opted to omit these data points. We also excluded a single outlier point, approximately 100\,\ms offset from the rest of the measurements for this target, likely due to moon contamination (99\% full at $36^\circ$ at the time of observation). This leaves us with the 286 RVs, measured between Dec 2021 and April 2025, with a median RV uncertainty of 0.92\,\ms.

\begin{figure*}
    \centering
    \includegraphics[width=0.98\textwidth]{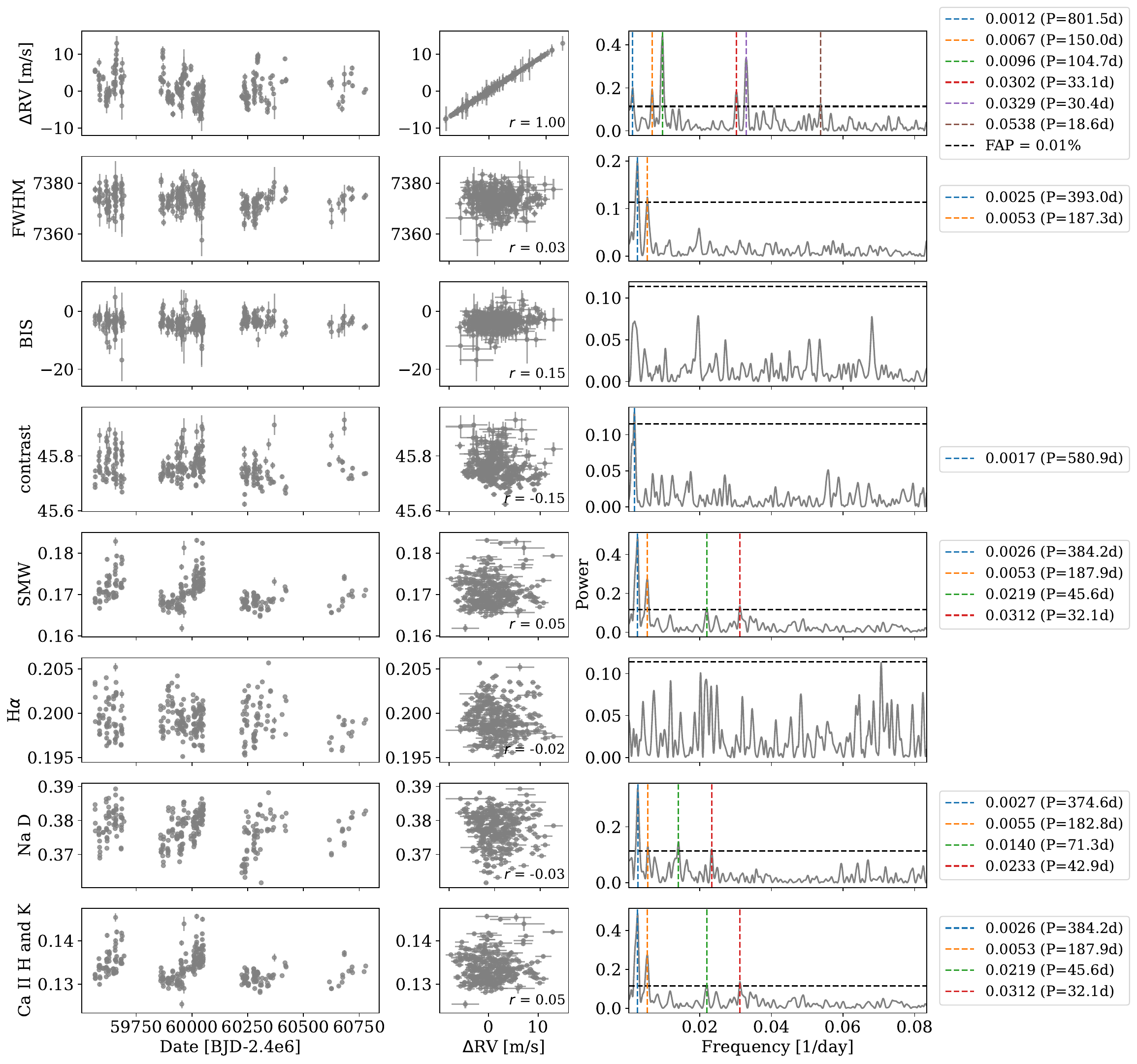}
    \caption{HARPS-N DRS 3.0.1 data products. The left column shows the time series of the radial velocities, followed by the various activity indicators. The middle column shows those values plotted with respect to the radial velocity. The Pearson r correlation coefficient is noted in the lower right of each subpanel. The right column shows the periodograms of these measurements. All of the significant periodicities (FAP$<0.0001$) are marked with vertical lines in all of the panels. We have excluded three contrast outliers, four SMW outliers, and four Ca II H and K outliers using 3-sigma clipping.}
    \label{fig:rv_act_inds_pgram}
\end{figure*}

We reduced these spectra with version 3.0.1 of the HARPS-N Data Reduction Software \citep[DRS,][]{dumusque_three_2021}. This pipeline outputs a number of data products, including the stellar radial velocity found using a G8 mask and stellar activity indicators: spectral bisector span (BIS), the full width at half maximum (FWHM) and contrast from the cross-correlation function (CCF) as well as H-alpha, Na D index, Ca II H and K index and Mount Wilson S-index (SMW), which we referenced in our analysis (Figure \ref{fig:rv_act_inds_pgram}).

\subsection{RV Periodograms}\label{subsect:RV_pgrams}

\begin{figure}
    \centering
    \includegraphics[width=0.99\linewidth]{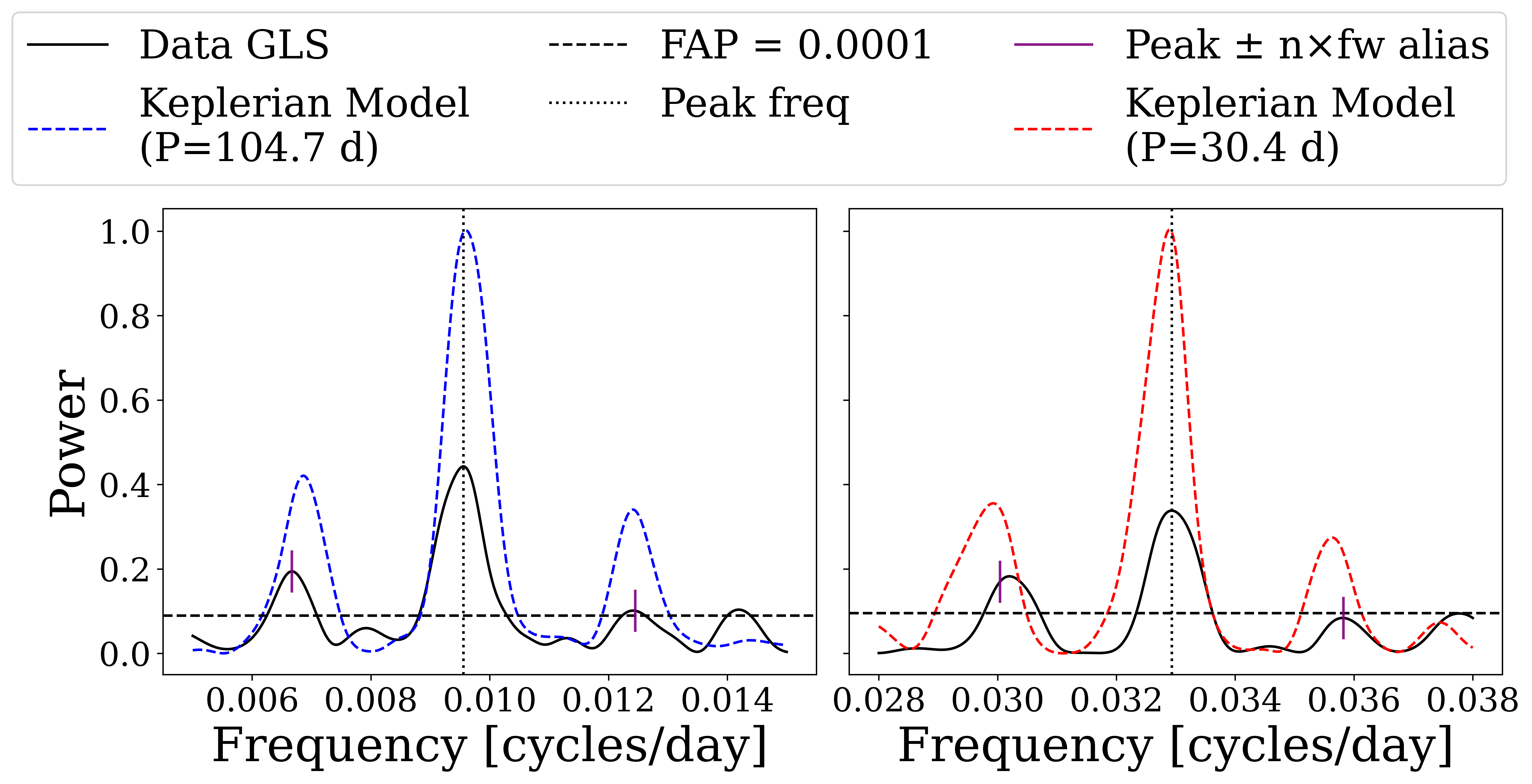}
    \caption{This figure demonstrates that two of the periodicities seen in the periodogram of the radial velocities are aliases of the 105\,d and 30\,d periodicities present in the RV dataset. Both panels show the periodogram of the RV dataset in black. The periodograms plotted in the dashed lines correspond to a Keplerian with that period, sampled at the times of the data. The short dashed vertical lines show the calculated alias of the main peak in the periodogram of the data and the main peak of the window function, which is at 346.06\,d.}
    \label{fig:aliases}
\end{figure}

We calculated periodograms of the RVs and stellar activity indicators using \texttt{Astropy}\footnote{https://www.astropy.org/}’s LombScargle function \citep[][Figure \ref{fig:rv_act_inds_pgram}]{the_astropy_collaboration_astropy_2018}. We found a number of periodicities above the 0.01\% False Alarm Probability (FAP) line, as calculated via the \cite{baluev_assessing_2008a} methodology: 18.6\,d, 30.4\,d, 33.1\,d, 104.7\,d, 150.0\,d, 801.5\,d. The 33.1\,d peak appears to be an alias of the 346.06\,d peak in the window function around the 30.4\,d periodicity, and the 150.0\,d peak appears to similarly be an alias of the 336\,d peak in the window function around the 104.7\,d periodicity (Figure \ref{fig:aliases}).

\begin{figure*}
    \centering
    \includegraphics[width=0.99\textwidth]{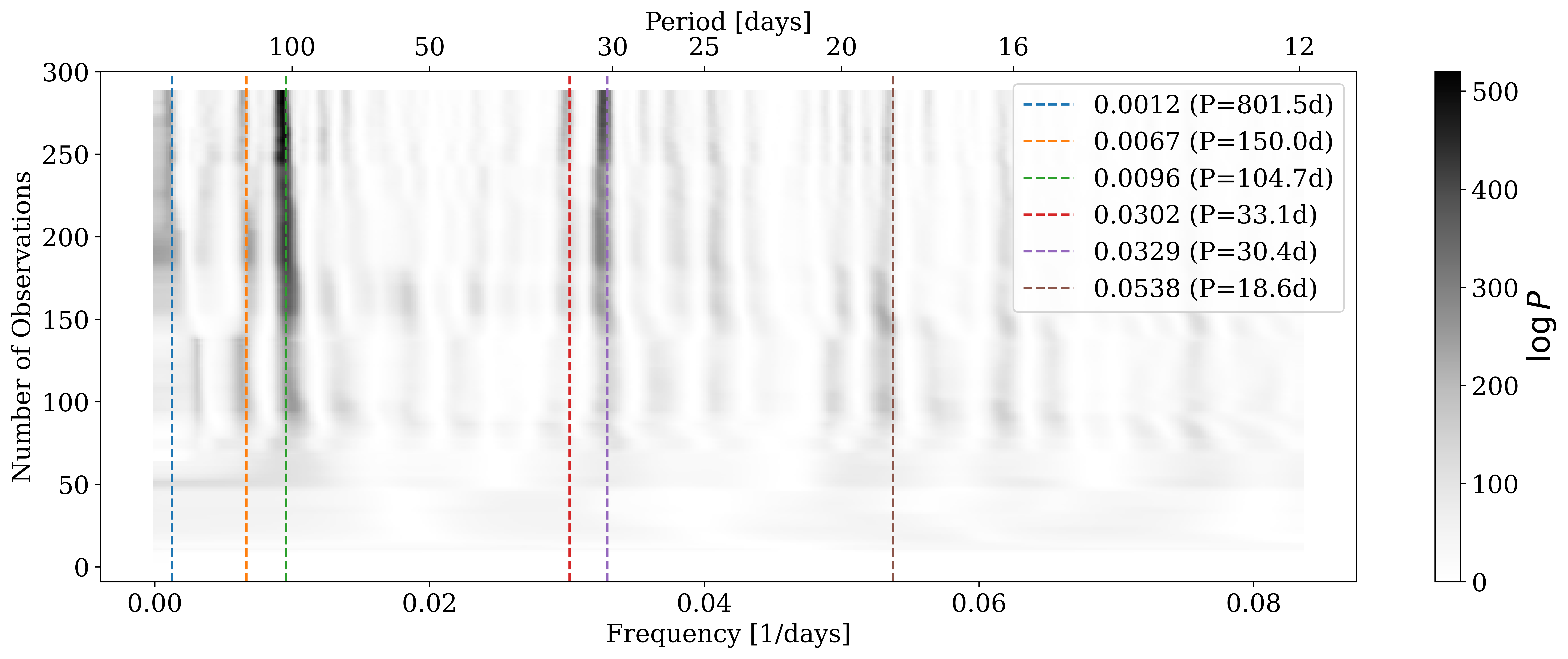}
    \caption{SBGLS for HD~60779's RV measurements. The dashed lines correspond to the statistically significant peaks in the classical Lomb-Scargle periodogram of the RVs (Figure \ref{fig:rv_act_inds_pgram}). The periodicities at 105\,d and 30\,d become more pronounced as more observations are added, while the 18.6\,d periodicity does not.}
    \label{fig:stacked_pgram}
\end{figure*}

We produced a stacked Bayesian General Lomb-Scargle periodogram (SBLGS, calculated via the procedure in \citealt{mortier_stacked_2017}). This shows that the 30.4\,d and 104.6\,d signals become stronger as we add more observations, as we would expect of a planet signal (Figure \ref{fig:stacked_pgram}). Stellar activity signals, by comparison, evolve over time and appear to shift or weaken as more observations are added, which is the case with the 18.6\,d periodicity. We expect the stellar rotation period to be no more than $19.15 \pm 3.29$\,d as dictated by the $\vsini$ measurement and stellar radius, but see no significant periodicity in the activity indicators in that range.

\subsection{RV Model Fitting}\label{subsect:RVmodelfitting}

To determine the origin of the different periodicities in the RV data, and to constrain the orbital period of HD~60779~b, we tried simultaneously fitting the data with various combinations of planet, stellar activity, and trend models. We performed these tests with \texttt{PyORBIT}\footnote{https://github.com/LucaMalavolta/PyORBIT} \citep{malavolta_pyorbit_2016, malavolta_ultrashort_2018}, which implements a Markov Chain Monte Carlo (MCMC) routine using \texttt{emcee}\footnote{https://github.com/dfm/emcee} \citep{foreman-mackey_emcee_2013}. For each planet, we fit an independent Keplerian model to the RVs, and to model stellar activity, we used a quasiperiodic Gaussian Process (GP).

\subsubsection{Model Comparison}

We first tested the hypothesis that the data can be explained by a single planet. To do this, we set a tight prior on $T_c$, restricting this value to within about 7 minutes, in accordance with the time of the observed TESS transit. With a wide, uninformative prior on $\Porb$ (12-200~d), the best-fit value is 107\,d. Such a Keplerian, however, leaves a strong 30d periodicity in the residuals. Restricting $\Porb<50$\,d, results in a best-fit $\Porb$ of 30\,d, with a 105\,d periodicity left in the residuals.

As stellar activity is a common source of periodicity in stellar RVs, we next tested how well a single Keplerian model, with the addition of a GP characterized by a dominant period $\Prot$, explained our RV data. First we bind the Keplerian with an uninformative uniform prior on the planet's $\Porb$ (12-200~d), while constraining the transit time in accordance with the observed transit. This model converged to a $\Porb=106.5$\,d and a $\Prot$ of 30\,d. This is suboptimal on two counts: first that best-fit stellar rotation period is in disagreement with the rotation period calculated based on the stellar radius and $\vsini$ (despite our providing this information as a Gaussian prior on $\Prot$, $19\pm3.5$). Second, the orbital period is now in conflict with the fact that no transit was observed in Sector 7 of the TESS data. We also attempted a fit where we constrained the orbital period of the transiting planet to 30\,d, but such a fit would not converge. Given that the RV data are not well explained by a single planet plus an activity model, we moved on to the next hypothesis, that there are additional planets in the system.

We next fit two Keplerian models simultaneously to the RV data. We constrained the orbital period of one of the Keplerians to $>50$\,d and the other to be $<50$\,d. We ran tests assuming that the outer planet is transiting and then that the inner planet is transiting (meaning that its phase is constrained by the observed transit.) When the outer planet was the transiting planet, we found best-fit orbital periods for the Keplerians of 107\,d and 30\,d, but a number of new significant periodic signals were left in the residuals of the fit. In the case where the inner planet was the transiting one, as we expect from our estimates of $\Porb$ (Section \ref{subsect:prelim_transit_analysis}), the best-fit orbital periods were 30\,d and 104\,d, but the 800\,d periodicity remained in the residuals.

\begin{figure*}
    \centering
    \includegraphics[width=0.99\textwidth]{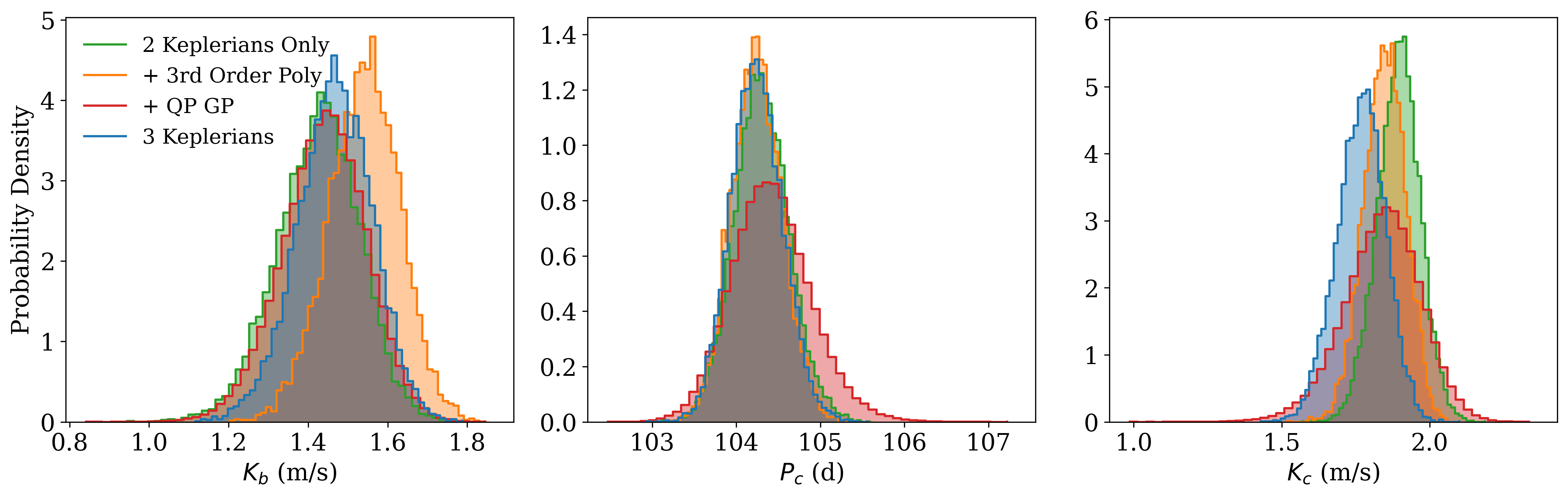}
    \caption{Posterior distributions for $K_b$, $P_c$ and $K_c$ given different RV model fits. The green histogram shows results when fitting the RVs with two Keplerians, the orange histogram shows results when fitting the RVs with two Keplerians plus a third-order polynomial, the red histogram shows results when fitting the RVs with two Keplerians plus a quasi-periodic GP, and the blue histogram shows results when fitting the RVs with three Keplerians. All of these fits were done fixing $T_b$ and $P_b$ in accordance with the observed transits. All of the results are consistent within $1\sigma$.}
    \label{fig:comp_models}
\end{figure*}

We tried modeling the long-period trend with three different functional forms. First, we tried adding a polynomial to the two Keplerians. A third-order polynomial is necessary to fit the 800\,d periodicity in our 1200\,d dataset (a second-order polynomial did not remove the 800\,d periodicity and a fourth-order polynomial resulted in a final coefficient consistent with zero) and resulted in a well-converged fit. Second, we added an additional Keplerian. This resulted in a well-converged fit, with the third keplerian's $\Porb=774_{-42}^{+48}$\,d, $K=1.13_{-0.21}^{+0.22}$\,\ms corresponding to $M_p=16.6^{+3.2}_{-3.1}$\,M$_\oplus$. Finally, we added a GP, with the $\Prot$ constrained by a Gaussian prior, in accordance with calculating a $\Prot$ from the $\vsini$ and stellar radius. All three of these choices result in model fits that explain all periodicity in the systems. These three model fits, as well as the model with just two Keplerians, all result in consistent parameters (within $1\sigma$ agreement, with comparable uncertainties) for the two inner planets, HD~60779~b and HD~60779~c (Figure \ref{fig:comp_models}).

Finally, we had to select our models for the final fitting, choosing between a two-Keplerian model, two Keplerians plus a third-order polynomial, two Keplerians plus a quasi-periodic GP, and a three-Keplerian model. When comparing the two-Keplerian model to all of the others using Akaike Information Criterion (AIC) and Bayesian Information Criterion (BIC) statistics, it was always disfavored. Regarding the quasi-periodic GP, while there are significant long-period periodicities in some of the activty indicators, none are at 800\,d. Regarding the potential third Keplerian, additional longer-baseline RVs or astrometry from Gaia DR4 would be necessary to confirm the presence of an outer planet in the system. Ultimately, we are not able to confirm nor rule out if this long-term periodicity is evidence of an outer planet, connected to stellar activity or a result of season-to-season instrumental offsets. In an effort to not mis-attribute the origin of this signal, we selected the third-order polynomial as the most agnostic option.
We again state that the best-fit parameters for HD~60779~b and HD~60779~c are consistent regardless of this model selection.

\subsubsection{Additional Reduction and Analysis}

In addition to fitting Keplerian and GP models to the HARPS-N RV measurements extracted using the DRS described above, we performed an additional, independent RV analyses on the same HARPS-N spectral dataset. The first step used the Self-Correlation Analysis of Line Profiles for Extracting Low-amplitude Shifts (\texttt{SCALPELS}). Described in detail in \cite{collier_cameron_separating_2021, john_impact_2022}, the \texttt{SCALPELS} algorithm calculates the cross-correlation function (CCF) resulting from correlating the observed spectra with a stellar template, and the autocorrelation function (ACF) of the CCF. It then uses the CCF and ACF to produce shape-driven differential RV measurements (largely the result of instrumental effects and stellar activity), and shift-driven differential RV measurements (largely the result of true dynamical Doppler shift). A \texttt{SCALPELS} analysis of the HD~60779 dataset showed that the RV variation does not project strongly on to the CCF shape basis vectors, suggesting that the RV variation is not significantly influenced by instrumental effects or stellar activity.

The second software used in this independent analysis is \texttt{kima}\footnote{https://github.com/j-faria/kima} \citep{faria_kima_2018}. This software fits an RV time series with Keplerian models as well as a quasi-periodic GP using the Diffusive Nested Sampling algorithm \citep{brewer_diffusive_2011} and can treat the number of Keplerians included in the fit as a free parameter. Using \texttt{kima} to fit the RV timeseries as calculated by \texttt{SCALPELS} leads to the conclusion that the 30\,d and 105\,d Keplerians are robustly detected, to greater than $10 \sigma$ significance with a false inclusion probability (FIP) $<2\%$. Stellar activity is found to be weakly present at a period of 20\,d, to $5.5\sigma$ significance with an FIP of 66.4\%. This analysis also found that the 30\,d planet candidate phase was consistent with the time of the observed transit and that the phase of the 105\,d planet candidate was not. Taken altogether, the analysis of the HARPS-N data using both the DRS with \texttt{PyORBIT}, as well as using \texttt{SCALPELS} with \texttt{kima} agree that the data are best explained by the presence of two planets in the system, as well as a weakly present additional signal, the source of which is unknown.

\section{Transit Follow-Up}\label{sect:transit_followup}

While we were able to constrain the orbital period of the transiting planet to P$=29.996\pm0.011$\,d using radial velocities and the first observed TESS transit, the uncertainty on the orital period was still prohibitively large for scheduling follow-up observations with facilities like the Hubble Space Telescope (HST) and the James Webb Space Telescope (JWST). Only the detection of an additional transit would constrain the orbital period well enough to enable follow-up transit observations.

\subsection{Predicting Transit Time}

In addition to the measurements of the orbital period we were able to make with our RV measurements, we were able to place a further constraint on the predicted transiting time by leveraging the earlier non-detection of the transit during TESS Sector 7. Using the best-fit orbital period for HD~60779~b, we found that an earlier transit of the planet would have fallen directly in the gap of TESS Sector 7. Propagating the full posteriors of the orbital period and time of central transit for HD~60779~b back in time to Sector 7 revealed that some of the $T_c$-$\Porb$ posterior were inconsistent with the non-detection of the earlier transit. This served as an additional constraint on the true $T_c$-$\Porb$ values for the inner planet.

We submitted a TESS DDT proposal (PI: Vanderburg) for high-cadence data of HD~60779 when the telescope revisited the target in Sector 88. Given the uncertainty in the transit time of HD~60779~b, however, we could not be certain that we would indeed observe an additional transit. 

\subsection{CHEOPS Data}

We proposed for and were awarded observations with the CHaracterising ExOPlanet Satellite (CHEOPS)\footnote{AO-5 PR250013, PI: DiTomasso}. CHEOPS is a 30cm space-based photometer primarily performing targeted exoplanet transit follow up observations \citep{benz_cheops_2021}. CHEOPS observed HD~60779 for 84.8 hours from UTC 12-16 February, 2025 with an integration time of 36\,s. These images were reduced and aperture photometry was performed using the CHEOPS DRP version 15.1.0 \citep{hoyer_expected_2020}. This pipeline performs bias, gain, non-linearity, dark current, flat field, bad pixel, smear and background corrections. We opted to use photometry extracted from a 22.5\,arcsec aperture. This particular aperture size balanced the influence of systematics (which increase with smaller apertures) with background noise (which increases with larger apertures). A transit matching the depth and duration of HD~60779 b was evident in the minimally processed DRP light curve.

Even though we were able to detect the transit in the standard DRP light curve, instrumental artifacts were clearly visible in the light curve. Therefore, we applied systematics corrections using the publicly available python package \texttt{PyCHEOPS}\footnote{https://github.com/pmaxted/pycheops} \citep{maxted_pycheops_2023}. We opted to consider only six hours before and after the time of central transit for our analysis. We detrended the flux of the transit light curve against the first- and second-order telescope rolling angle, denoted in \texttt{PyCHEOPS} as \texttt{dfdsinphi}, \texttt{dfdcosphi}, and \texttt{dfdsin2phi}. We selected these parameters by testing all of the available detrending parameters in \texttt{PyCHEOPS}, and selecting those that decreased the RMS residual without increasing the AIC and BIC statistics. We also considered the Bayes factors, which compare the relative evidence for models with and without a given decorrelation basis vector. Our chosen parameters removed the effect of additional background (largely reflected light from the Moon and Earth) that comes in and out of the field of view as the telescope rolls during the observations. We fit these detrending parameters simultaneously with a transit model. The resulting transit depth and duration from this fit were consistent with the TESS transit.

\begin{figure*}
    \centering
    \includegraphics[width=0.99\textwidth]{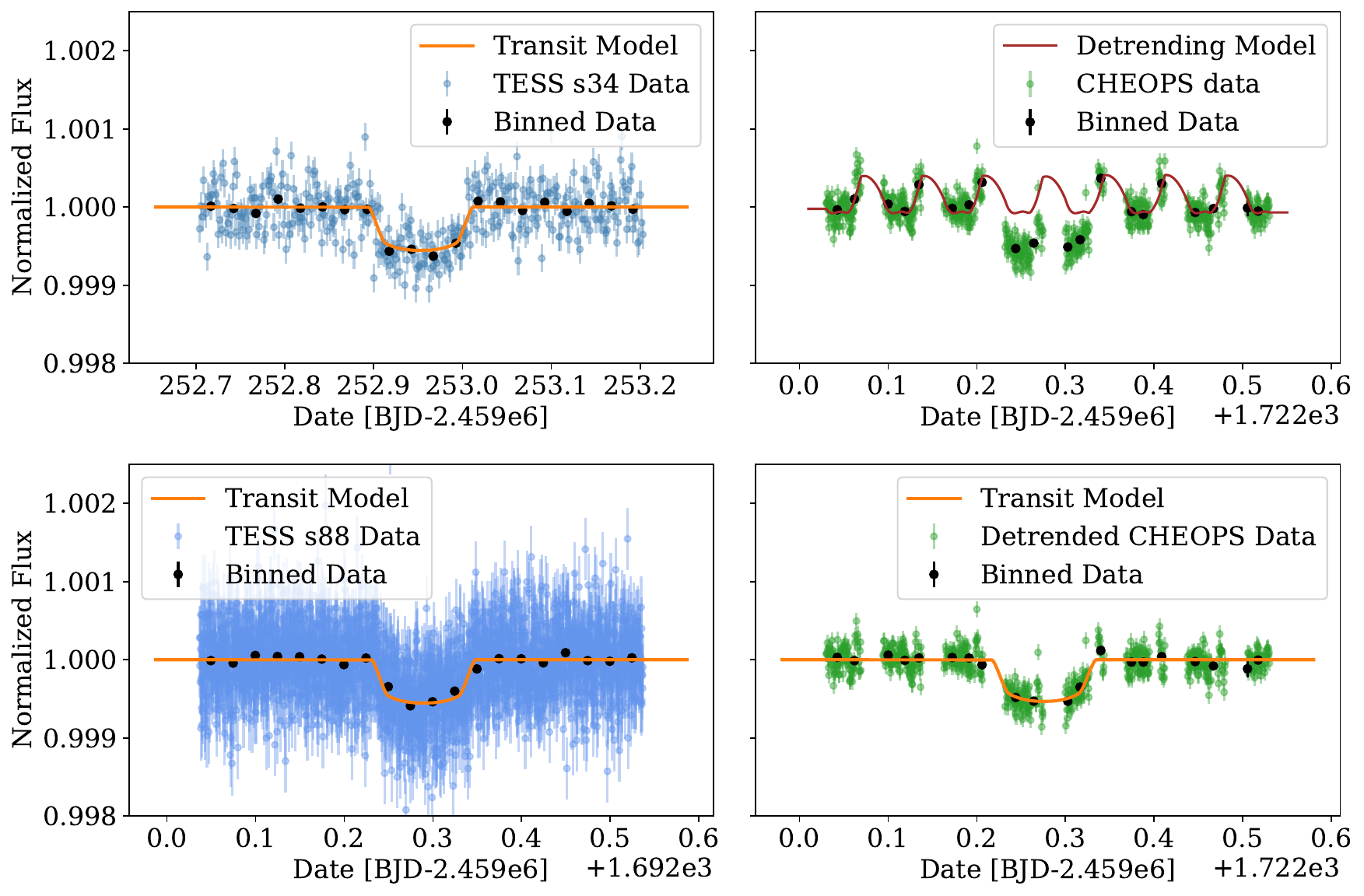}
    \caption{Transit light curves and fits for HD~60779~b. The transit models were all fit simultaneously with the RVs, details listed in Table \ref{table:gp&kep}. The upper left panel shows the flattened 120\,s cadence data from TESS Sector 34. This is the original transit detection. The lower left panel shows the flattened 20\,s cadence data from TESS Sector 88. The upper right panel shows the CHEOPS light curve, with the detrending model plotted. The lower right panel shows the detrended CHEOPS transit light curve, fit with the transit model. In all panels, the black points show the data in bins of 0.025\,d.}
    \label{fig:transits}
\end{figure*}

We then removed the best-fit detrending model from the DRP light curve. In doing so, we calculated point-by-point uncertainties for the model via error propogation, using the uncertainties on the individual detrending parameters. We added these uncertainties in quadrature to the DRP-reported flux uncertainties when subtracting the detrending model from the DRP light curve, so we could then fit the detrended CHEOPS light curve simultaneously with the TESS light curves and RVs using \texttt{PyORBIT} (Figure \ref{fig:transits}).

\subsection{Additional TESS Transit}

TESS observed HD~60779 during Sector 88 (Jan 14-Feb 11, 2025). TESS captured a complete transit of HD~60779~b in this sector, bringing the total number of observed transits of the inner planet to three. We include the 20\,s cadence SPOC data of this transit when fitting for the final parameters of HD~60779~b, as described in Section \ref{sect:final_RV_transit_analysis}.

\section{Stellar Characterization}\label{sect:starchar}

HD~60779 is a G1 V star \citep{houk_michigan_1999} located $35.393_{-0.022}^{+0.020}$\,pc away \citep{bailer-jones_estimating_2021}. As a relatively bright sunlike star, HD~60779 has been characterized in multiple previous papers, including \cite{masana_effective_2006, holmberg_geneva-copenhagen_2009, pace_chromospheric_2013, mortier_functional_2013, zenoviene_stellar_2015, fuhrmann_ancient_2021, hardegree-ullman_bioverse_2023} prior to the detection of its planets. Adding to this, we used a number of the datasets presented in this paper to perform our own stellar analysis.

\subsection{Stellar Parameters}

We used several approaches to characterize HD~60779, which we ultimately averaged to reduce the impacts of systematics. To the HARPS-N spectra, we applied three spectroscopy-based stellar characterization methods: Stellar Parameter Classification \citep[SPC,][]{buchhave_abundance_2012,buchhave_three_2014}, ARES+MOOG \citep{sousa_ares_2014}, and CCFpams \citep{malavolta_atmospheric_2017}. SPC uses a template matching approach to determine stellar effective temperature, $\log g$, $v\sin i$ and [m/H] metallicity, defined as the fraction of elements heavier than hydrogen and helium present in a star’s outer atmosphere. The ARES + MOOG algorithm uses an equivalent width approach to measure $T_{\rm eff}$, $\log g$, microturbulence, and [Fe/H] metallicity. \texttt{CCFpams}\footnote{https://github.com/LucaMalavolta/CCFpams} leverages cross-correlations of the observed spectra with reference templates, built to include particular spectral features that are known to be sensitive to different photospheric parameters such as $T_{\rm eff}$ and $\log g$.

In order to determine the physical stellar parameters, particularly mass, radius and age, from the spectroscopic values we derived from each method, we also used stellar characterization methods that leverage available survey photometry, Gaia parallax measurements and isochrones. We considered the photospheric parameters from each of the three above methods and, following \cite{mortier_k2-111_2020}, fit the target to both MESA isochrones and Stellar Tracks \citep[MIST, ][]{dotter_mesa_2016} and Dartmouth Stellar Evolution Database models \citep{dotter_dartmouth_2008, mortier_bgls_2015,mortier_k2-111_2020}.

For our final approach to stellar characterization, we used the code \texttt{uberMS}\footnote{\url{https://github.com/pacargile/uberMS/}} \citep{ting_payne_2019,cargile_minesweeper_2020}. This method uses neural networks to interpolate between Kurucz stellar models, simultaneously fitting stellar photometry and spectroscopy to determine stellar parameters. It includes the option to simultaneously fit MIST isochrone models as well \citep{choi_mesa_2016}. For this method, we used a spectrum of HD~60779 from the Tillinghast Reflector Echelle Spectrograph (TRES) spectrograph. TRES is a high-resolution spectrograph on the 1.5\,m Tillinghast Telescope located at the Fred Lawrence Whipple Observatory at Mt. Hopkins, AZ \citep{szentgyorgyi_precision_}. This spectrum was obtained on 02 Feb 2025 with a single exposure of 600\,s. We applied \texttt{uberMS}, both with and without isochrone fitting, to the spectrum of this star from the TRES spectrograph to simultaneously determine $\teff$, $\log{g}$, $\mathrm{[Fe/H]}$, $\mathrm{[\alpha/Fe]}$, $\rstar$ and $\mstar$. For these measurements, we followed the methodology of, and adopted the recommended uncertainties from, \cite{pass_metallicities_2025}.

\begin{figure*}
    \centering
    \includegraphics[width=0.99\textwidth]{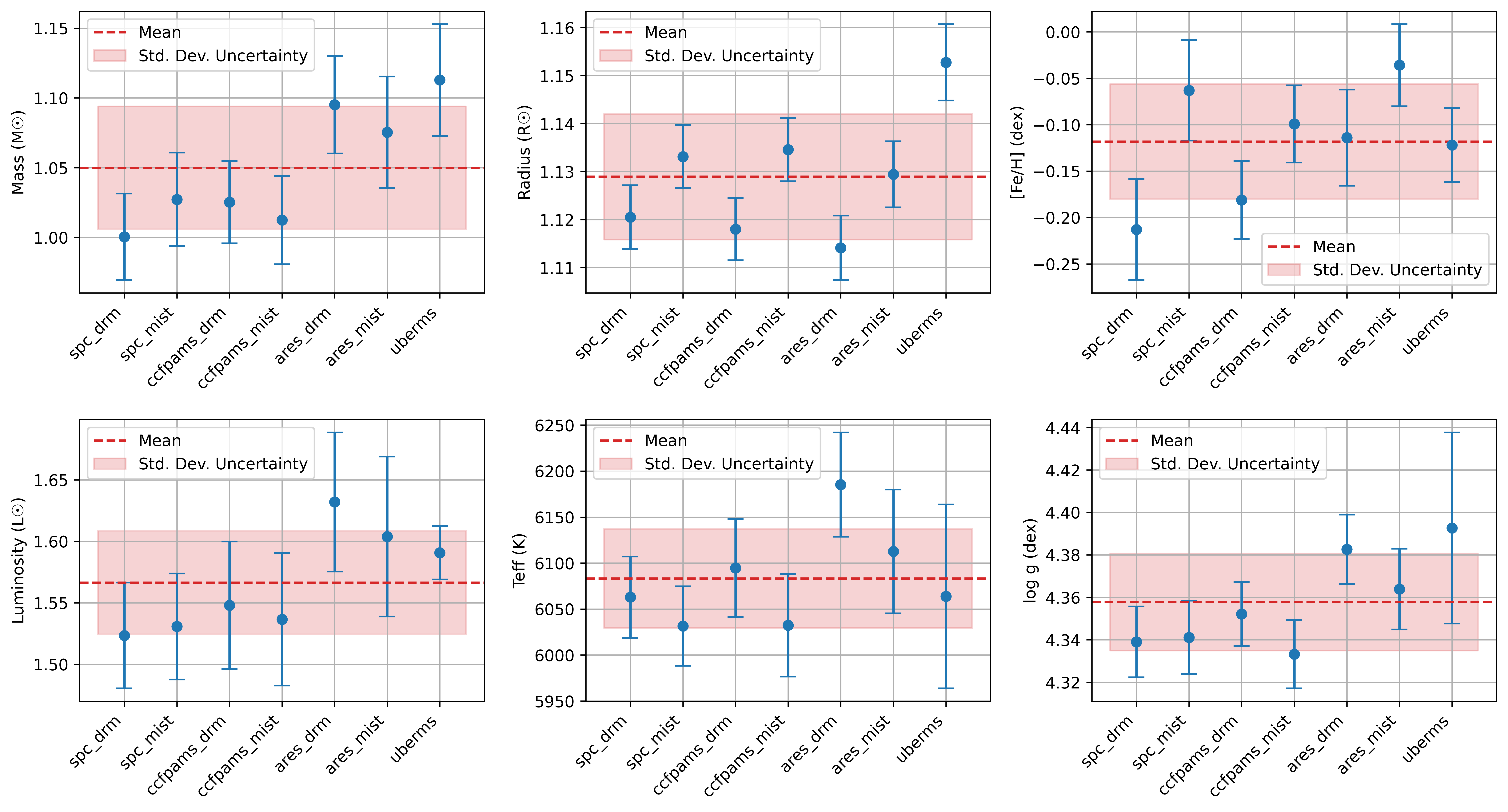}
    \caption{Estimates for the stellar parameters from the different methods described in Section \ref{sect:starchar}. The red dashed line denotes the average value and the red shaded region denotes the adopted uncertainty.}
    \label{fig:stellar_params}
\end{figure*}

Ultimately, we combined the results that leveraged spectroscopy, photometry and isochrones to characterize HD~60779. We assigned a characteristic uncertainty to each parameter, informed by the standard deviation between the best-fit values from each method. We note that this standard deviation, as well as the uncertainties on the measurements from each method, were all comparable. All of these results are plotted in Figure \ref{fig:stellar_params} and our final values are listed in Table \ref{table:results_star}.

\begin{table}[]
    \caption{HD~60779 stellar parameters. Values are from this work, unless otherwise noted.}
\centering
\begin{tabular}{ll}
\smallskip
\\

\multicolumn{1}{c}{Stellar Properties}                               & \multicolumn{1}{c}{Value} \\ 

\noalign{\smallskip}
\hline
\noalign{\smallskip}
Alternate names & GJ 9237 \\
 & 	TIC 65672998\\
RA (HMS) & 07h 36m 01.43s $^a$ \\
Dec (DMS) & -03 09 06.38 $^a$ \\
$G$ (mag) & 7.023 $\pm$ 0.028 $^a$ \\
$\teff$ (K) & $6081 \pm 60$  \\
$[$Fe/H$]$ (dex) & $-0.12 \pm 0.05$ \\
$[\alpha/\text{Fe}]$ (dex) & $0.08 \pm 0.03$ \\
log($g$) (cgs) & $4.356 \pm 0.021$ \\
$\mstar$ (M$_\odot$) & $1.050 \pm 0.044$ \\
$\rstar$ (R$_\sun$) & $1.129 \pm 0.013$ \\
L$_{\text{bol}}$ (erg s$^{-1}$) & $1.566 \pm 0.042$ \\
$v \sin{i_\star}$ (km s$^{-1}$) & $3.0 \pm 0.5$ \\
$\Prot$$_{\text{,max}}$ (d) & $19\pm3.5$ \\
RV (\kms) & $129.75\pm0.12$ $^a$ \\
Parallax (mas) & $28.221 \pm 0.019$ $^a$\\
Distance (pc) & $35.393_{-0.022}^{+0.020}$ $^b$ \\
SpT & G1 V $^c$ \\
\\
\end{tabular}

\footnotesize{$^a$ \cite{gaia_collaboration_gaia_2023}, $^b$ \cite{bailer-jones_estimating_2021}, $^c$ \cite{houk_michigan_1999} }

\label{table:results_star}

\end{table}

\subsection{Galactic Dynamics}

The barycentric RV of HD~60779 is $129.75 \pm 0.12$\,\kms \citep{gaia_collaboration_gaia_2023}. In this respect, HD~60779 stands out among bright ($G<14$), nearby ($<40$\,pc) stars as well as among exoplanet hosts for its remarkably high systemic radial velocity.

The thin and thick disks of the Milky Way are two populations of stars with distinct ages and chemistry, and are often identified by their galactic kinematics. Thick disk stars are older, have lower [Fe/H] and higher [$\alpha$/Fe], and tend to be kinematically hotter than their thin disk counterparts \citep[][and the references therein]{haywood_accurate_2018}. Despite its high barycentric RV, the full UVW kinematics of HD~60779 are not fully consistent with thick disk membership. Using the galactic population membership relationship from \cite{bensby_exploring_2014}, local number densities of the thin and thick disk from \cite{fantin_canadafrance_2019}, the local standard of rest from \cite{schonrich_local_2010}, and \texttt{Astropy} to translate Gaia DR3 RA, Dec and proper motion to galactic coordinates, yield a thick disk / thin disk membership probability of 4.9, which corresponds to an indeterminate thick disk membership \citep{medina_galactic_2022}. Using the \texttt{Gala}\footnote{https://github.com/adrn/gala} code \citep{price-whelan_gala_2017} we integrated HD~60779’s orbit to determine the maximum height above the Galactic plane it is expected to reach. For HD~60779, this value is 73\,pc, putting it well within the range of typical thin disk stars \citep{li_evolution_2017}. Additionally, our [Fe/H] and [$\alpha$/Fe] measurements of -0.12\,dex and 0.08\,dex (Section \ref{sect:starchar}) are well-consistent with the galactic thin disk \citep{soubiran_abundance_2005}. Although HD~60779’s high systemic RV could suggest thick disk membership, the evidence instead supports its classification as a thin disk star.

\begin{figure*}
    \centering
    \includegraphics[width=0.99\textwidth]{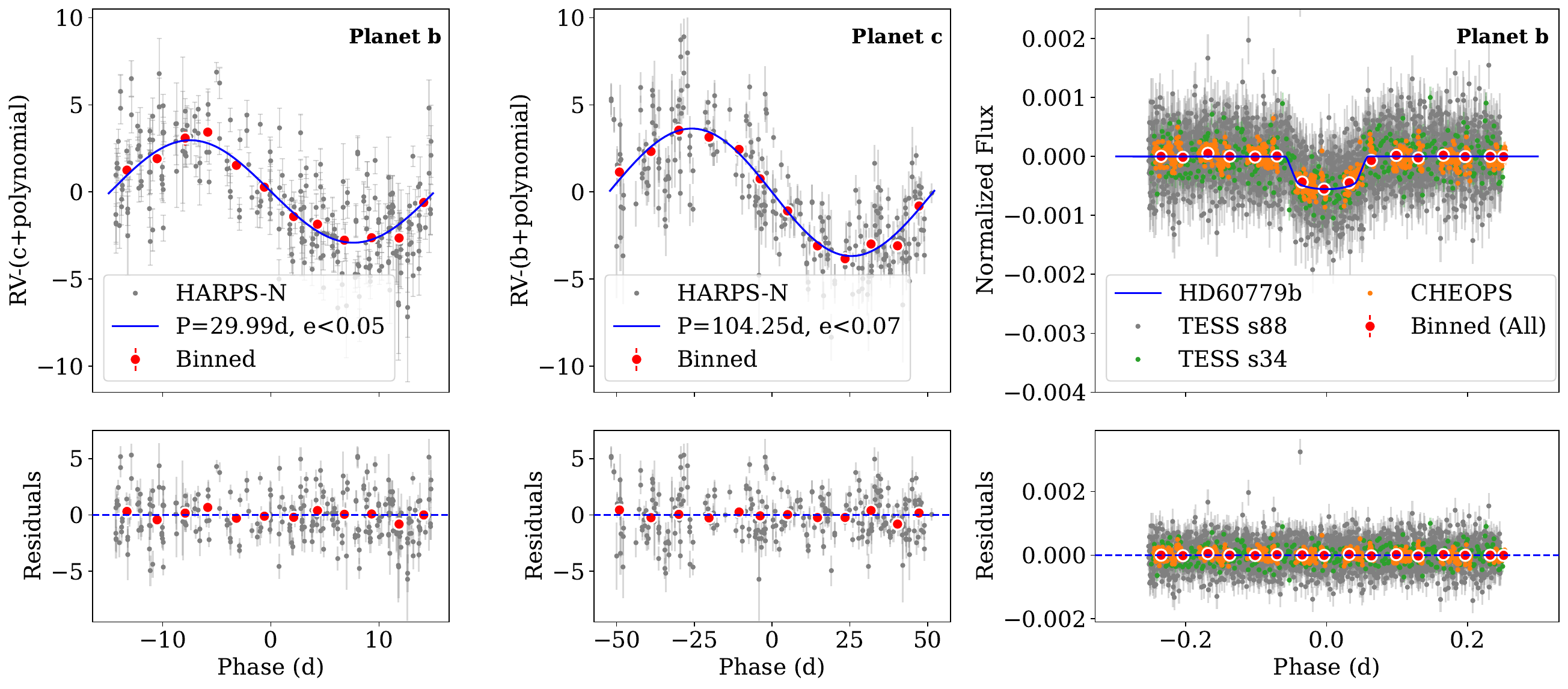}
    \caption{Final RV and transit fits. The left and center panels show the phase-folded RV curves for HD~60779~b and HD~60779~c, respectively. The right panel shows the phase-folded transit light curves from TESS Sectors 34 and 88, and from CHEOPS. The red points in all panels show binned values. The stated eccentricities are the $1\sigma$ upper limits.}
    \label{fig:final_fit}
\end{figure*}

\begin{figure}
    \centering
    \includegraphics[width=0.95\linewidth]{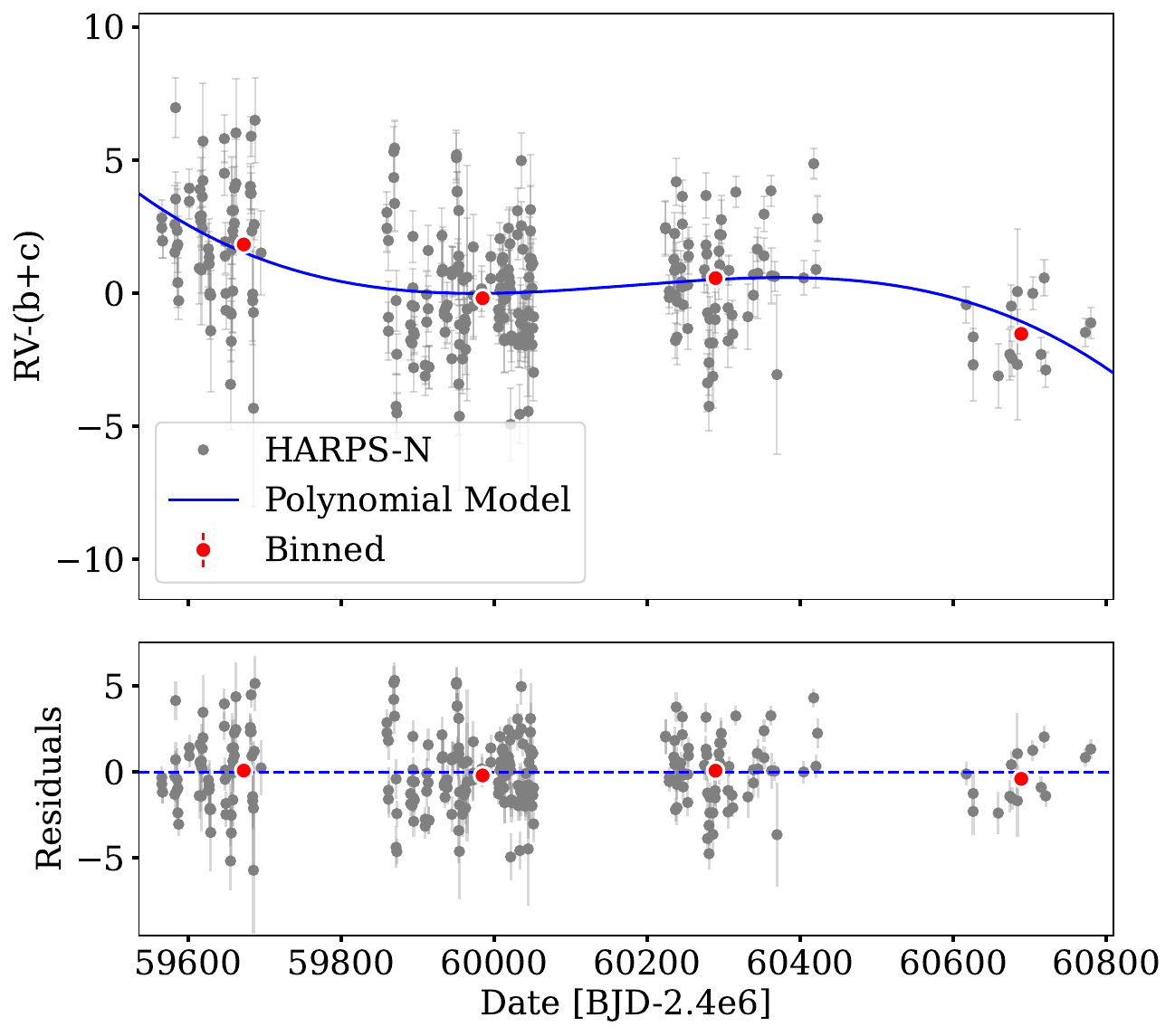}
    \caption{Third-order polynomial as fit simultaneously to the RVs in the final RV and transit fits.}
    \label{fig:polyfit}
\end{figure}

\section{Joint RV and Transit Analysis}\label{sect:final_RV_transit_analysis}

To determine our final best-fit parameters for the planets in this system, we simultaneously fit the available HARPS-N radial velocities, two TESS transits and the CHEOPS transit with two Keplerian models (one transiting, one non-transiting) and a third-order polynomial fit to the RVs using \texttt{PyORBIT}.
For each planet, we fit an independent Keplerian model to the RVs, parameterized with respect to orbital period (P), time of conjunction (Tc), RV semi-amplitude (K), eccentricity (e) and argument of pericenter ($\omega$).
We parameterized $e$ and $\omega$ as $\sqrt{e}\sin \omega$ and $\sqrt{e}\cos\omega$ following \cite{eastman_exofast_2013}.
We constrained the period of the inner planet to be 25-50\,days and we constrained the period of the outer planet to 50-200\,days (see Section \ref{subsect:RVmodelfitting}). 

We fit the transit of the inner planet simultaneously with the three light curves: the SPOC reduction of the 120\,s cadence data from TESS Sector 34, the SPOC reduction of the 20\,s cadence data from TESS Sector 88 and the detrended CHEOPS data.
For all photometric datasets, we fit 0.25 days on either side of midtransit. We further detrended the TESS data using \texttt{keplerspline}\footnote{https://github.com/avanderburg/keplerspline}, fitting a spline to 0.25 days on either side of midtransit while masking out the transit itself. We used the \texttt{choosekeplerspline} function to optimize the break-point spacing based on the Bayesian Information Criterion, resulting in a break-point spacing of 0.5\,d for both sectors.
We used \texttt{batman} \citep{kreidberg_batman_2015} to model the transit of the inner planet, and parameterized the limb-darkening as prescribed in \citep{kipping_efficient_2013}. We used \texttt{LDTK}\footnote{https://github.com/hpparvi/ldtk} \citep{parviainen_ldtk_2015,husser_new_2013} to estimate the quadratic law coefficients for the TESS and CHEOPS lightcurves, and adopted those as the center of our Gaussian priors, with a width of 0.1.

\begin{figure*}
    \centering
    \includegraphics[width=0.99\textwidth]{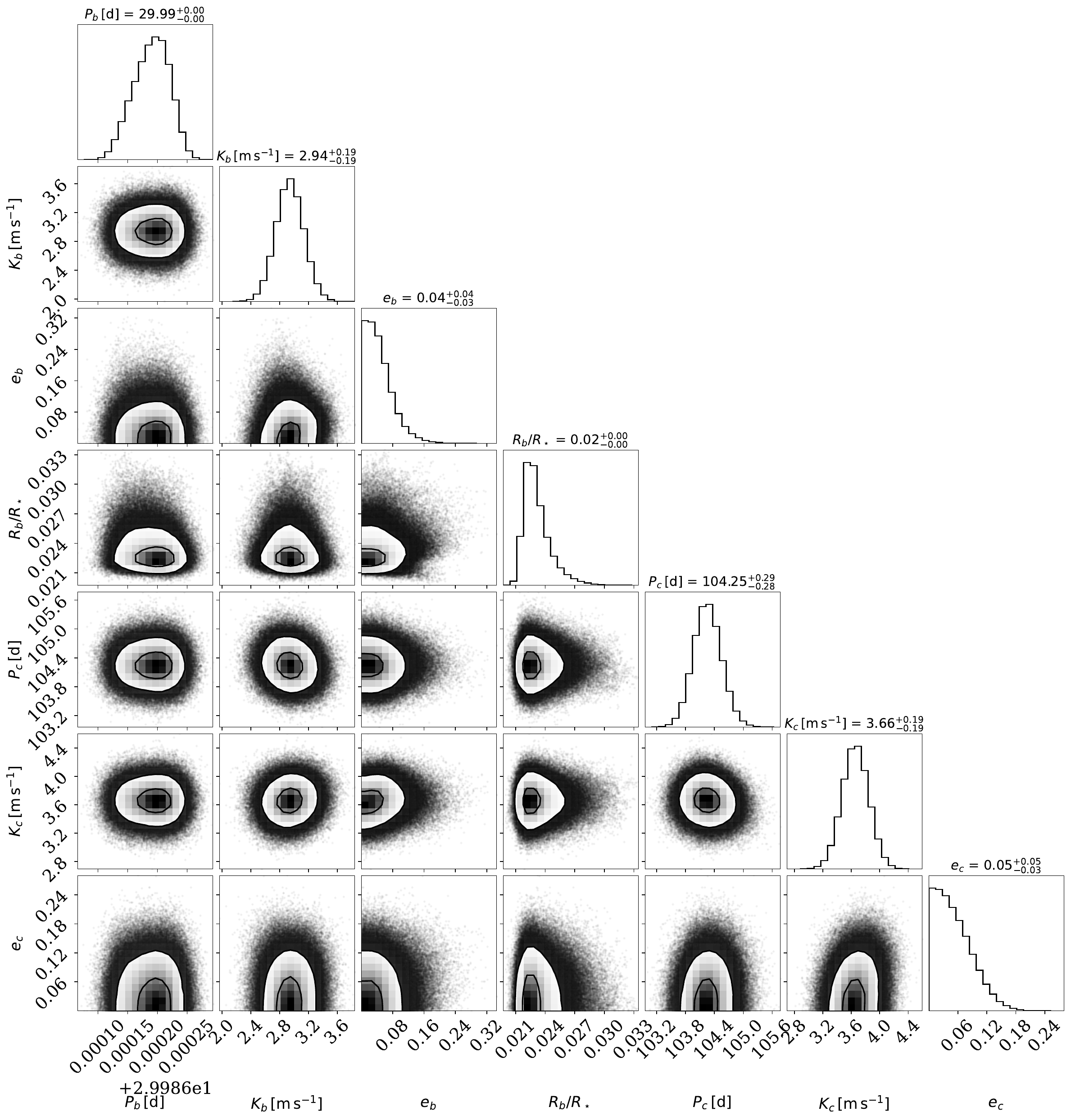}
    \caption{Corner plot of the planet parameter posteriors.}
    \label{fig:corners}
\end{figure*}

We ran the fitter for 250,000 steps, discarding the first 38,000 steps as burn-in and used a thinning factor of 100. From the statistics on the posterior of the sampler parameters, we determine our best fit values and one-sigma uncertainties. Our final Keplerian and transit fits are shown in Figure \ref{fig:final_fit}, final third-order polynomial RV trend fit is shown in Figure \ref{fig:polyfit}, cornerplots for the planet parameters are plotted in Figure \ref{fig:corners} and all priors and posteriors are listed in Table \ref{table:gp&kep}.

\begin{table*}
    
    \caption{Priors and posteriors final fit.}\label{table:gp&kep}
\centering

\begin{tabular}{lll}
\toprule
\multicolumn{1}{c}{Parameter} & \multicolumn{1}{c}{Prior} &
\multicolumn{1}{c}{Posterior} \\
\midrule

Planet b, Model Parameters & & \\
\noalign{\smallskip}
\hline
\noalign{\smallskip}
Tc (BJD-2.45e6) &  $\mathcal{U}(9252.9511,9252.9557)$  & $9252.9531^{+0.0014}_{-0.0013}$ \\
$R_p/\rstar$ &  $\mathcal{U}(0,0.5)$  & $0.02638^{+0.00078}_{-0.00073}$ \\
$P$ (d) & $\mathcal{U}(25,50)$  & $29.986175^{+0.000030}_{-0.000033}$ \\
$K$ (\ms) & $\mathcal{L}_2\mathcal{U}(-9.9658,10.9658)$ & $2.94_{-0.19}^{+0.20}$ \\
$b$ &  $\mathcal{U}(0,2)$  & $0.9230^{+0.0078}_{-0.0081}$ \\
$\sqrt{e}\sin{\omega}$ & $\mathcal{U}(-1,1)$ & $-0.04\pm0.17$ \\
$\sqrt{e}\cos{\omega}$ & $\mathcal{U}(-1,1)$ & $0.09_{-0.11}^{+0.10}$ \\
\noalign{\smallskip}
\hline
\noalign{\smallskip}
Planet b, Derived Parameters & & \\
\noalign{\smallskip}
\hline
\noalign{\smallskip}
R$_p$ (R$_\oplus$) &  & $3.250^{+0.100}_{-0.098}$ \\
$M_p (M_\oplus)$ & & $14.7^{+1.1}_{-1.0}$ \\
$a$ (AU) & & $0.1920^{+0.0026}_{-0.0027}$ \\
$i$ ($^\circ$) &  & $88.566^{+0.055}_{-0.051}$ \\
$e$ & & $<0.053$ (1$\sigma$)\\
Transit Duration, $T_{14}$ (hr) & & $2.810^{+0.091}_{-0.101}$ \\
$\rho$ (\gcm) & & $2.35\pm0.28$ \\
$\teq$ (K) & & $497 \pm 38$ \\
TSM & & $71\pm10$ \\
\noalign{\smallskip}
\hline
\noalign{\smallskip}
Planet c, Model Parameters & & \\
\noalign{\smallskip}
\hline
\noalign{\smallskip}
$T_c$ (BJD-2.45e6) &  $\mathcal{U}(9200, 9300)$  & $9272.6\pm2.4$ \\
$P$ (d) & $\mathcal{U}(50,200)$  & $104.25^{+0.30}_{-0.29}$ \\
$K$ (\ms) & $\mathcal{L}_2\mathcal{U}(-9.9658,10.9658)$ & $3.66\pm0.19$ \\
$\sqrt{e}\sin{\omega}$ & $\mathcal{U}(-1,1)$ & $0.11^{+0.14}_{-0.17}$ \\
$\sqrt{e}\cos{\omega}$ & $\mathcal{U}(-1,1)$ & $-0.05_{-0.15}^{+0.16}$ \\
\noalign{\smallskip}
\hline
\noalign{\smallskip}
Planet c, Derived Parameters & & \\
\noalign{\smallskip}
\hline
\noalign{\smallskip}
$M_p \sin{i}$ $(M_\oplus)$ & & $27.7\pm1.6$ \\
$a$ (AU) & & $0.4406^{+0.0061}_{-0.0063}$ \\
$e$ & & $<0.067$ (1$\sigma$)\\
\noalign{\smallskip}
\hline
\noalign{\smallskip}
Stellar Parameters from Transit Fit \\
\noalign{\smallskip}
\hline
\noalign{\smallskip}
$\rho_\star$ $(\mathrm{\rho_\odot})$ & $\mathcal{N} (0.730\pm0.035)$ & $0.733\pm0.035$ \\
$u_1,_\text{TESS}$ & $\mathcal{N} (0.38\pm0.1)$ & $0.427_{-0.085}^{+0.083}$ \\
$u_2,_\text{TESS}$ & $\mathcal{N} (0.16\pm0.1)$ & $0.207_{-0.087}^{+0.088}$ \\
$u_1,_\text{CHEOPS}$ & $\mathcal{N} (0.49\pm0.1)$ & $0.511_{-0.079}^{+0.078}$ \\
$u_2,_\text{CHEOPS}$ & $\mathcal{N} (0.16\pm0.1)$ & $0.189_{-0.084}^{+0.083}$ \\
\noalign{\smallskip}
\hline
\noalign{\smallskip}

Polynomial Parameters & & \\
\noalign{\smallskip}
\hline
\noalign{\smallskip}
$C_3$ & $\mathcal{U}(-10^6, 10^6)$ & $(-1.86\pm0.39)\times10^{-8}$ \\
$C_2$ & $\mathcal{U}(-10^6, 10^6)$ & $(9.4\pm1.8)\times10^{-6}$ \\
$C_1$ & $\mathcal{U}(-10^6, 10^6)$ & $(6.3_{-8.7}^{+8.6})\times10^{-4}$ \\
T$_\text{0}$ (BJD-2.45e6) & Fixed & 10006.790917 \\

\noalign{\smallskip}
\hline
\noalign{\smallskip}

Data Offset and Jitter & & \\
\noalign{\smallskip}
\hline
\noalign{\smallskip}
RV offset (\ms) & $\mathcal{U} (119603.8162,139624.3471)$ & $129612.27_{-0.17}^{+0.18}$ \\
RV jitter (\ms) & $\mathcal{U} (0.0045,419.8611)$ & $1.75\pm0.10$ \\
TESS Sect 34 jitter (ppm) & $\mathcal{U} (0,18100)$ & $211^{+14}_{-13}$ \\
TESS Sect 88 jitter (ppm) & $\mathcal{U} (0,41000)$ & $153^{+18}_{-20}$ \\
CHEOPS jitter (ppm) & $\mathcal{U} (0,10600)$ & $100\pm6$ \\
\bottomrule
\end{tabular}

\end{table*}

\section{Conclusions and Discussion}\label{sect:conclusions}

HD~60779 is a bright ($V=7.2$), nearby (35~pc),  Sun-like ($\mstar=1.050\pm 0.044M_\odot$, $\rstar=1.129\pm0.013R_\odot$) star. It hosts two planets, HD~60779~b ($P_\text{orb}=29.986175^{+0.000030}_{-0.000033}$d, $R_p=3.250^{+0.100}_{-0.098} R_\oplus$, $M_p=14.7^{+1.1}_{-1.0} M_\oplus$) and HD~60779~c ($P_\text{orb}=104.25^{+0.30}_{-0.29}$d, $\mp\sini=27.7 \pm 1.6 M_\oplus$). HD~60779~b has a bulk density of $2.35\pm0.28$\,\gcm. According to the composition curves presented in \cite{zeng_growth_2019}, HD~60779~b is consistent with a 100\% H$_2$O composition or a 2\% H$_2$ envelope with a rocky core.

Planet c, if assumed to have the same inclination as the inner planet, would not transit. Even if planet c were to have an inclination closer to 90$^{\circ}$ than planet b, resulting in it transiting the host star, there would be no overlap between the existing photometric coverage and the predicted transit time for the outer planet, as constrained by the RVs. In conclusion, we have no evidence that HD~60779~c is transiting, but we cannot rule it out.

Both of these planets' orbits are consistent with being circular despite their long periods.
Following the equations of \cite{adams_longterm_2006}, the circularization timescales for these two planets are on the order of $10^{13}-10^{15}$\,years, depending on the choice of tidal circularization factor. This incredibly long circularization timescale indicates that they originated on low-eccentricity orbits and have not undergone strong dynamical interactions since.

\begin{figure}
    \centering
    \includegraphics[width=0.99\linewidth]{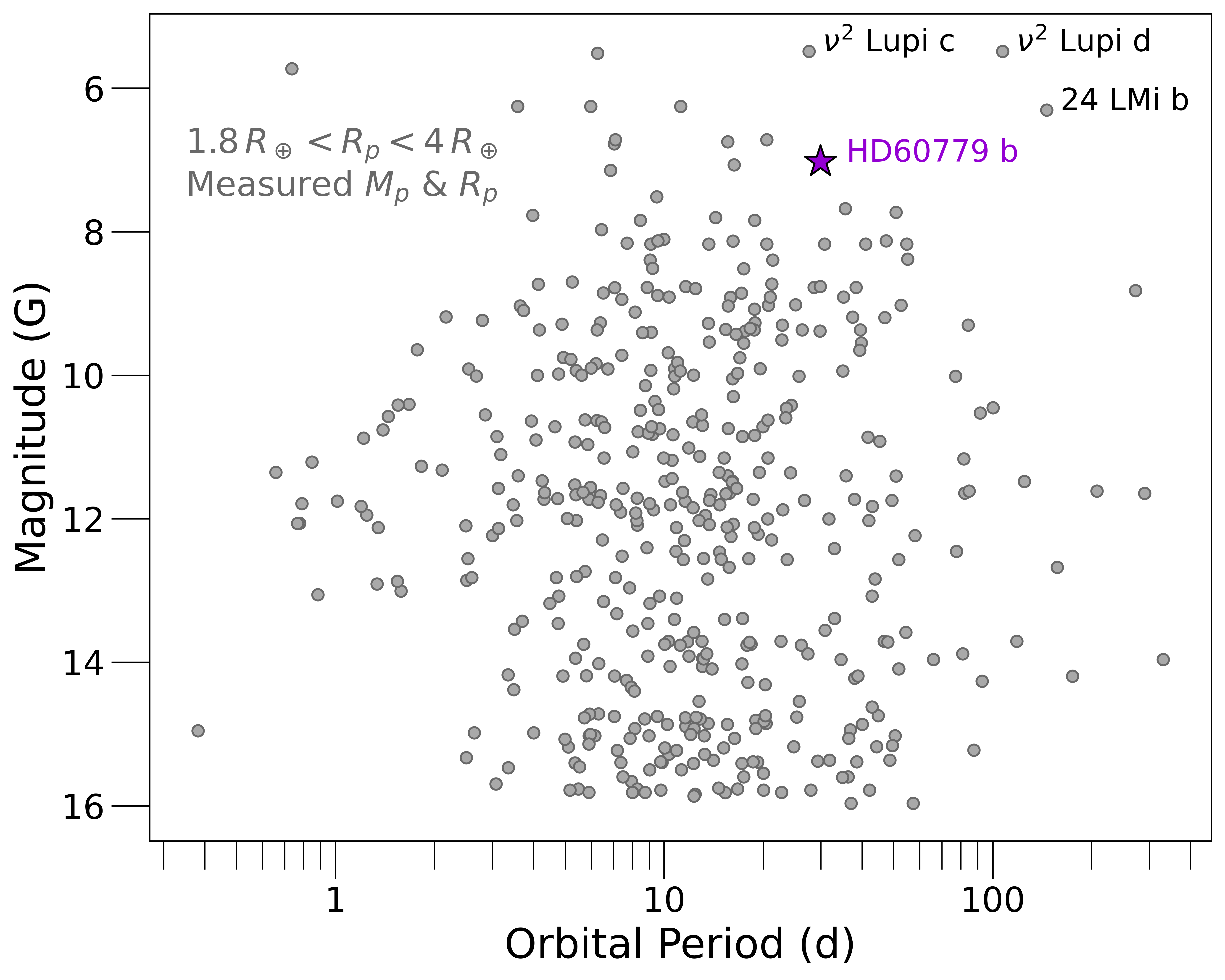}
    \caption{Host star G magnitudes and orbital periods of their sub-Neptunes ($2 R_\oplus < R_p < 4R_\oplus$) with measured masses and radii. We have labeled the only other planets on orbits longer than 25d around stars brighter than $G=7.5$, along with HD~60779~b.}
    \label{fig:incontext_subNeptunes}
\end{figure}

\begin{figure}
    \centering
    \includegraphics[width=0.99\linewidth]{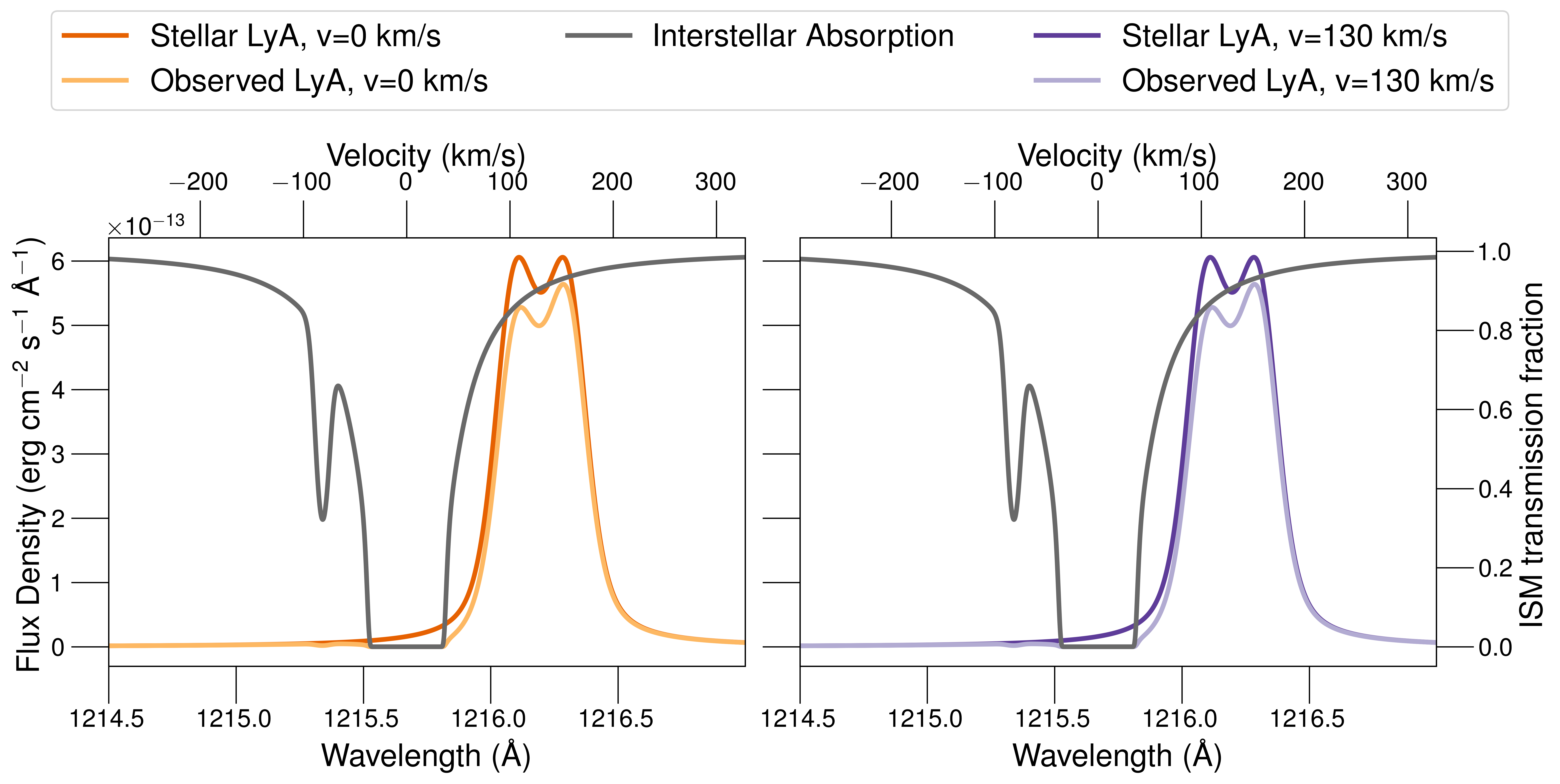}
    \caption{Simulated stellar Ly$\alpha$ emission and corresponding absorption by the interstellar medium for a star with a systemic RV of 0\,\kms (upper panel) and with that of HD~60779, 130\,\kms (lower panel). A systemic RV of 130\,\kms would Doppler shift stellar Ly$\alpha$ emission out of the interstellar absorption band, giving us access to the entire Ly$\alpha$ feature. These plots were generated using the \texttt{Lyapy} code from \cite{youngblood_muscles_2016}.}
    \label{fig:lya}
\end{figure}

\begin{figure}
    \centering
    \includegraphics[width=0.99\linewidth]{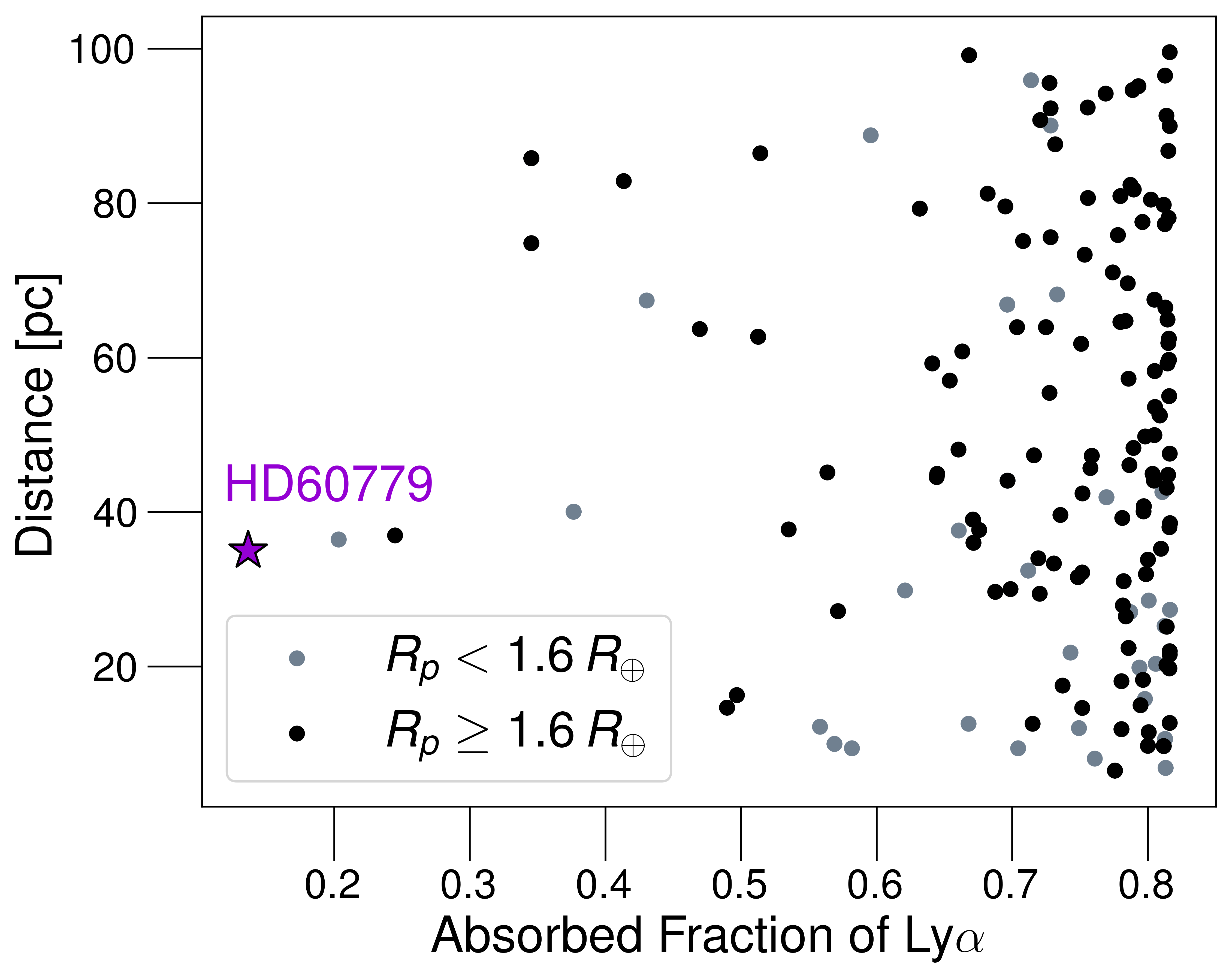}
    \caption{Exoplanets with measured masses and radii, and with systemic radial velocities available in the Exoplanet Archive. The plot shows the distance to each host star versus the fraction of emitted Ly$\alpha$ that we expect to be absorbed by the ISM. We calculated this by computing the ratio of the integrated flux from an emitted Ly$\alpha$ line to the integrated flux after accounting for absorption by the ISM, and subtracting this ratio from one. Note that we do not estimate the intrinsic strength of the Ly$\alpha$ line for any host star. We plot this metric against distance because nearer systems typically suffer less ISM absorption, as their light travels through a shorter column of interstellar material. Planet size is also indicated, since we do not expect atmospheric escape for planets smaller than 1.6 $R_\oplus$. HD~60779~b stands out as both nearby and expected to experience little Ly$\alpha$ ISM absorption.}
    \label{fig:incontext_lya}
\end{figure}

HD~60779~b is one of the longer-period sub-Neptune-sized planets to be discovered, particularly around bright stars (Figure \ref{fig:incontext_subNeptunes}), and it is accessible to valuable follow-up observations. It is an especially promising target for Lyman-$\alpha$ (Ly$\alpha$) transit observations using HST. The 121~nm Ly$\alpha$ line is absorbed by neutral hydrogen and has been used to detect atmospheric escape on Neptune-sized exoplanets \citep{ehrenreich_giant_2015}. There is a fundamental challenge to observing Ly$\alpha$ absorption from exoplanet atmospheres, which is that neutral hydrogen in the interstellar medium also absorbs Ly$\alpha$. As a result, exoplanetary Ly$\alpha$ detections rely on observations of the high-velocity wings of the stellar Ly$\alpha$ in nearby targets \citep{linsky_lymana_2014}. However, the high systemic RV of HD~60779 would Doppler shift any stellar Ly$\alpha$ emission, and any accompanying absorption from HD~60779~b’s atmosphere, out of the interstellar absorption band (Figure \ref{fig:lya}). 
According to the NASA Exoplanet Archive, HD~60779 has the highest systemic RV of any star hosting a planet with a measured mass and radius, making it uniquely suited for Ly$\alpha$ observations (Figure \ref{fig:incontext_lya}). An initial search for stellar Ly$\alpha$ emission in HD~60779 will soon be conducted (HST GO 17804), and if stellar Ly$\alpha$ emission is detected, then a transit observation of planet b may provide the first constraints on exospheric hydrogen in the low-velocity core of the Ly$\alpha$ line.

Space-based transmission spectroscopy may also be useful for observing HD~60779~b’s atmosphere. HD~60779~b has a transmission spectroscopy metric \citep[TSM,][]{kempton_framework_2018} of 71, putting it among the highest 12 when compared to other long period ($\Porb>25$\,d) sub-Neptunes with measured masses and radii on the exoplanet archive. Instrument filters and modes, however, will need to be carefully selected prior to observations with JWST, as HD~60779 is brighter than the typical atmospheric target ($J=6.1$).

Using the constraint on secondary eclipse probability for transiting exoplanets given in \cite{von_braun_constraints_2010}, accounting for the posterior distributions of the best-fit orbital parameters of HD~60779~b, the probability of secondary eclipse is $\geq 97\%$. HD~60779~b’s emission spectroscopy metric \citep[ESM,][]{kempton_framework_2018} is 19, putting it among the highest 7 when compared to other long period ($\Porb>25$\,d) sub-Neptunes with measured masses and radii on the exoplanet archive.

HD~60779~b is also a promising candidate for ground-based transit follow-up, particularly for measuring the Rossiter-McLaughlin (RM) effect. High-precision RV measurements taken during a planet’s transit can reveal the sky-projected angle between the planet’s orbital axis and its host star’s spin axis. This alignment or misalignment provides clues into the dynamical formation and evolution of a planetary system. Few RM measurements have been made for small, longer-period ($R_p<4R_\oplus$, P$>$10d) planets thus far, and already they reveal a wide range of alignments \citep{huber_stellar_2013,wang_stellar_2018}. If aligned, we expect HD~60779~b to have an RM amplitude of $1.3\ms$. Although challenging, similar RM amplitudes have been successfully measured, e.g. \cite{dalal_nearly_2019, polanski_aligned_2025}.

Opportunities to observe HD~60779~b transit from the ground are rare due to its long orbital period. A partial transit is expected to be visible from Hawaii in Dec 2025 (one or more hours of pre-transit baseline and 78\% of transit). The next full transit will be visible from both Hawaii (with less than an hour of pre-transit baseline and more than an hour post-transit) and Arizona (more than an hour of pre-transit baseline and less than an hour post-transit) in Dec 2026, followed by two more transits visible from Hawaii in January and February 2027 (both with an hour or more of pre- and post-transit baseline). The soonest transit that is $>80\%$ accessible from Chile will be in Nov 2027 (with very little out of transit baseline). The following two transits (Dec 2027 and Jan 2028) will also be visible from Chile with significantly more pre-transit baseline. The soonest transit that is $>80\%$ accessible from La Palma will occur in Nov of 2028 (with an hour or more pre- and post-transit baseline). Despite these limited observing windows, HD~60779~b is a promising candidate to help build our understanding of planetary system architectures.




\section*{Acknowledgements}
This material is based upon work supported by the National Science Foundation Graduate Research Fellowship under Grant No. DGE1745303 and by NASA under award number 80GSFC21M0002.
This publication is funded in part by the Alfred P. Sloan Foundation under grant G202114194.

This paper includes data collected by the TESS
mission, which are publicly available from the Mikulski Archive for Space Telescopes (MAST). Funding for the TESS mission is provided by NASA’s Science Mission Directorate. We acknowledge the use of public TESS data from pipelines at the TESS Science Office and at the TESS Science Processing Operations Center. Resources supporting this work were provided by the NASA High-End Computing (HEC) Program through the NASA Advanced Supercomputing (NAS) Division at Ames Research Center for the production of the SPOC data products.

This work is based on observations made with the Italian Telescopio Nazionale Galileo (TNG) operated on the island of La Palma by the Fundación Galileo Galilei of the INAF (Istituto Nazionale di Astrofisica) at the Spanish Observatorio del Roque de los Muchachos of the Instituto de Astrofisica de Canarias. The HARPS-N project was funded by the Prodex Program of the Swiss Space Office (SSO), the Harvard University Origin of Life Initiative (HUOLI), the Scottish Universities Physics Alliance (SUPA), the University of Geneva, the Smithsonian Astrophysical Observatory (SAO), the Italian National Astrophysical Institute (INAF), University of St. Andrews, Queen’s University Belfast, and University of Edinburgh. The HARPS-N Instrument Project was partially funded through the Swiss ESA-PRODEX Programme. Based on observations made with the Italian Telescopio Nazionale Galileo (TNG) operated on the island of La Palma by the Fundación Galileo Galilei of the INAF (Istituto Nazionale di Astrofisica) at the Spanish Observatorio del Roque de los Muchachos of the Instituto de Astrofisica de Canarias

This study uses CHEOPS data observed as part of the Guest Observers (GO) programmes CH\_PR0013 (PI DiTomasso). CHEOPS is an ESA mission in partnership with Switzerland with important contributions to the payload and the ground segment from Austria, Belgium, France, Germany, Hungary, Italy, Portugal, Spain, Sweden, and the United Kingdom. The CHEOPS Consortium would like to gratefully acknowledge the support received by all the agencies, offices, universities, and industries involved. Their flexibility and willingness to explore new approaches were essential to the success of this mission. CHEOPS data analysed in this article will be made available in the CHEOPS mission archive (\url{https://cheops.unige.ch/archive_browser/}).

FPE would like to acknowledge the Swiss National Science Foundation (SNSF) for supporting research with HARPS-N through the SNSF grants nr. 140649, 152721, 166227, 184618 and 215190. MPi acknowledges support from the European Union – NextGenerationEU (PRIN MUR 2022 20229R43BH) and the ``Programma di Ricerca Fondamentale INAF 2023''.
AM, AAJ, and BSL acknowledge funding from a UKRI Future Leader Fellowship, grant number MR/X033244/1. AM and FR acknowledge a UK Science and Technology Facilities Council (STFC) small grant ST/Y002334/1. ACC acknowledges support from STFC consolidated grant number ST/V000861/1 and UKRI/ERC Synergy Grant EP/Z000181/1 (REVEAL). FR is supported by the Science and Technology Funding Council ST/Y002334/1.


\clearpage

\bibliographystyle{aa_url}
\bibliography{ref}







\clearpage



\clearpage






\end{document}